\documentclass[isre,nonblindrev]{informs3} % current default for manuscript submission
\DoubleSpacedXI % current default line spacing

\usepackage{natbib}
 \bibpunct[, ]{(}{)}{,}{a}{}{,}%
 \def\bibfont{\small}%
\TheoremsNumberedThrough     % Preferred (Theorem 1, Lemma 1, Theorem 2)
\EquationsNumberedThrough    % Default: (1), (2), ...
\usepackage{subcaption}
\usepackage{pgfplots}
\pgfplotsset{compat=1.16}
\usepackage{booktabs}
\usepackage{pifont}
\newcommand{\cmark}{\ding{51}}
\newcommand{\xmark}{\ding{55}}
\usepackage{multirow}
\usepackage{multicol}

\renewcommand{\theARTICLETOP}{}

\begin{document}
%%%%%%%%%%%%%%%%

%\setcounter{page}{1}
%\FIRSAPAGE{000}%
%\LASAPAGE{000}%
%\LONGFIRSAPAGE{0001} %
%\DOI{10.1287/xxxx.0000.0000}%


\RUNTITLE{A Comparison of Methods for Treatment Assignment}

% \TITLE{Bundling Information Goods of Decreasing Value}
\TITLE{A Comparison of Methods for Treatment Assignment with an Application to Playlist Generation}

\ARTICLEAUTHORS{%
\AUTHOR{Carlos Fern\'{a}ndez-Lor\'{i}a}
\AFF{Hong Kong University of Science and Technology}
\AUTHOR{Foster Provost}
\AFF{New York University}
\AUTHOR{Jesse Anderton}
\AFF{Spotify, New York, USA}
\AUTHOR{Benjamin Carterette}
\AFF{Spotify, New York, USA}
\AUTHOR{Praveen Chandar}
\AFF{Spotify, New York, USA}
} % end of the block

\ABSTRACT{%
This study presents a systematic comparison of methods for individual treatment assignment, a general problem that arises in many applications and has received significant attention from economists, computer scientists, and social scientists. We group the various methods proposed in the literature into three general classes of algorithms (or metalearners): learning models to predict outcomes (the O-learner), learning models to predict causal effects (the E-learner), and learning models to predict optimal treatment assignments (the A-learner). We compare the metalearners in terms of (1) their level of generality and (2) the objective function they use to learn models from data; we then discuss the implications that these characteristics have for modeling and decision making. Notably, we demonstrate analytically and empirically that optimizing for the prediction of outcomes or causal effects is not the same as optimizing for treatment assignments, suggesting that in general the A-learner should lead to better treatment assignments than the other metalearners. We demonstrate the practical implications of our findings in the context of choosing, for each user, the best algorithm for playlist generation in order to optimize engagement. This is the first comparison of the three different metalearners on a real-world application at scale (based on more than half a billion individual treatment assignments). In addition to supporting our analytical findings, the results show how large A/B tests can provide substantial value for learning treatment assignment policies, rather than simply choosing the variant that performs best on average.}%

%\KEYWORDS{deterministic inventory theory; infinite linear programming duality; 
%  existence of optimal policies; semi-Markov decision process; cyclic schedule}

\HISTORY{}

\maketitle
%%%%%%%%%%%%%%%%%%%%%%%%%%%%%%%%%%%%%%%%%%%%%%%%%%%%%%%%%%%%%%%%%%%%%%

% Samples of sectioning (and labeling) in ISRE
% NOTE: (1) \section and \subsection do NOT end with a period
%       (2) \subsubsection and lower need end punctuation
%       (3) capitalization is as shown (title style).
%
%\section{Introduction.}\label{intro} %%1.
%\subsection{Duality and the Classical EOQ Problem.}\label{class-EOQ} %% 1.1.
%\subsection{Outline.}\label{outline1} %% 1.2.
%\subsubsection{Cyclic Schedules for the General Deterministic SMDP.}
%  \label{cyclic-schedules} %% 1.2.1
%\section{Problem Description.}\label{problemdescription} %% 2.
\vspace{-30pt}
\section{Introduction}

Systems that make automated decisions are often deployed with the underlying goal of improving (rather than just predicting) outcomes. For example, when targeting online ads or retention incentives, the goal is often to encourage customers to make a purchase or stay with the company rather than just predicting their behavior. This type of task is more generally known as a treatment assignment problem~\citep{manski2004statistical}, where each possible course of action corresponds to a different `treatment' (e.g., `show ad' vs `do not show ad'), and ideally each individual is assigned to the treatment associated with the most beneficial outcome (e.g., the one with the highest profit). 

Treatment assignment policies may be estimated from data using statistical modeling, allowing decision makers to map individuals to the best treatment according to their characteristics (e.g., preferences, behaviors, history with the firm). However, there are different ways in which one could proceed, and various methods for the estimation of treatment assignment policies from sample data have been proposed across different fields, including econometrics~\citep{manski2004statistical}, data mining~\citep{lo2002true}, and multi-armed bandits~\citep{beygelzimer2009offset}. 

This paper gathers these various methods into three general classes of algorithms (or metalearners) for individualized treatment assignment, all of which have been proposed in the literature. Table~\ref{tab:metalearners} describes these metalearners in terms of their estimand (i.e., the model that each metalearner intends to estimate) and their resulting treatment assignment policy.\footnote{When the goal is to identify the treatments that lead to the lowest outcomes (e.g., when $Y$ is the number of hospitalizations), the arg max should be replaced with an arg min in this and all subsequent equations.} The first metalearner (\textbf{Outcome Learner, O-learner}) learns a model that predicts the expected outcome ($Y$) given individual-level characteristics ($X=x$) and a specific treatment assignment ($T=j$). Each individual is then assigned to the treatment with the best (e.g., largest) predicted outcome. The second metalearner (\textbf{Effect Learner, E-learner}) learns a model for heterogeneous causal effect estimation and assigns each individual to the treatment with the largest predicted causal effect. Finally, the third metalearner (\textbf{Assignment Learner, A-learner}) directly learns and assigns individuals to the treatments that are expected to have the best outcome. 

\begin{table}
    \renewcommand{\arraystretch}{0.7} 
    \centering
    \begin{tabular}{c|c|c}
        \toprule
        \multirow{2}{*}{\textbf{Metalearners}}& \textbf{Estimand in the} & \textbf{Estimated treatment}\\
        & \textbf{learning procedure} & \textbf{assignment policy}\\
        \midrule
        $\begin{matrix} \text{Outcome Learner} \\ \text{(O-learner)} \end{matrix}$ 
         &$\mu(x, j) = \mathbb{E}[Y|X=x,T=j]$&$\argmax\limits_j ~\hat{\mu}(x,j)$\\
         \hline
        $\begin{matrix} \text{Effect Learner} \\ \text{(E-learner)} \end{matrix}$ 
         &$\tau(x, j) = \mu(x, j) - \mu(x, 0)$&$\argmax\limits_j ~\hat{\tau}(x,j)$\\
         \hline
        $\begin{matrix} \text{Assignment Learner} \\ \text{(A-learner)} \end{matrix}$ 
         &$a^*(x) = \argmax\limits_j ~\tau(x,j)$& $\hat{a}^*(x)$\\
         \hline
         \multicolumn{3}{l}{\footnotesize $Y$ is the outcome of interest, $X$ are individual-level characteristics, and $T$ is the treatment assignment.}\\
         \bottomrule
    \end{tabular}
    \caption{Metalearners for estimating treatment assignment policies}
    \label{tab:metalearners}
\end{table}

At a first glance, the policies estimated by the metalearners in Table~\ref{tab:metalearners} may look the same: they all seek to assign the treatment with the best outcome. As a main contribution, we reveal two key distinctions between these learning approaches and the implications that these distinctions have for personalized treatment assignment. 

The first distinction between metalearners is the level of generality of the tasks that the machine-learned models can perform. For instance, models that predict outcomes ($\hat{\mu}$ in Table~\ref{tab:metalearners}) may be used to estimate causal effects ($\hat{\tau}$ in Table~\ref{tab:metalearners}), whereas models that predict causal effects generally cannot predict outcomes. Thus, a more general metalearner may be preferable when the machine-learned models are important for other reasons besides identifying the treatment assignment with the largest outcome, such as when there are decision-making constraints that depend on predicted outcomes or causal effects. 

The second distinction is in the objective function that each metalearner optimizes when learning models. The first metalearner optimizes for outcome prediction, the second for causal effect prediction, and the third for optimal action (treatment assignment) prediction. It is clear from Table~\ref{tab:metalearners} that when models are highly accurate (i.e., $\hat{\mu}=\mu$, $\hat{\tau}=\tau$, and $\hat{a}^*=a^*$), then all the metalearners produce the same policy. In practice, however, machine-learned models are derived from sample data, so each metalearner may lead to a different treatment assignment policy depending on how its objective function frames the prediction problem. Importantly, and as we discuss in detail in this paper, optimizing models to predict outcomes or causal effects is not the same as optimizing models to predict treatment assignments. 

\begin{figure*}
    \centering
    \begin{subfigure}{.49\textwidth}
    \centering
     \begin{tikzpicture}[scale=0.8]
        \linespread{0.5}
        \begin{axis}[
            %hide y axis,
            ylabel = {Outcome},
            ymax=5,
            ymin=0,
            xmax= 2.5,
            xmin = 0.5,
            xticklabels={$T=1$ (better), $T=2$ (worse)}, xtick={1,2},
            axis y line*=left,
            axis x line*=bottom]
        \addplot[only marks, mark=*,red] plot coordinates {
            (1, 4.7)
            (2, 0.3)};
        \addlegendentry{~Predictions ($\hat{\mu}$)}
        \addplot[only marks, mark=*,blue] plot coordinates {
            (1, 3)
            (2, 2)};
        \addlegendentry{~True values ($\mu$)}
        \addplot[no marks,dashed] plot coordinates {
            (1, 4.7)
            (1, 3)};
        \addplot[no marks,dashed] plot coordinates {
            (2, 2)
            (2, 0.3)};
        \addlegendentry{~Prediction error}
        \node[align=center] at (axis cs:1,1.2) {Assigned \\ Treatment};
        \addplot[style={thick},->] coordinates {(1,0.7) (1,0.3)};
    \end{axis}
    \end{tikzpicture}
    \caption{Model with worse $\hat{\mu}$}
    \label{fig:outcome-high}
    \end{subfigure}
    \begin{subfigure}{.49\textwidth}
    \centering
     \begin{tikzpicture}[scale=0.8]
        \linespread{0.5}
        \begin{axis}[
            %hide y axis,
            ylabel = {Outcome},
            ymax=5,
            ymin=0,
            xmax= 2.5,
            xmin = 0.5,
            xticklabels={$T=1$ (better), $T=2$ (worse)}, xtick={1,2},
            axis y line*=left,
            axis x line*=bottom]
        \addplot[only marks, mark=*,red] plot coordinates {
            (1, 2.2)
            (2, 2.8)};
        \addlegendentry{~Predictions ($\hat{\mu}$)}
        \addplot[only marks, mark=*,blue] plot coordinates {
            (1, 3)
            (2, 2)};
        \addlegendentry{~True values ($\mu$)}
        \addplot[no marks,dashed] plot coordinates {
            (2, 2.8)
            (2, 2)};
        \addplot[no marks,dashed] plot coordinates {
            (1, 3)
            (1, 2.2)};
        \addlegendentry{~Prediction error}
        \node[align=center] at (axis cs:2,1.2) {Assigned \\ Treatment};
        \addplot[style={thick},->] coordinates {(2,0.7) (2,0.3)};
    \end{axis}
    \end{tikzpicture}
    \caption[size=10]{Model with better $\hat{\mu}$}
    \label{fig:outcome-low}
    \end{subfigure}
    
    \caption{\textbf{Comparison of outcome prediction vs treatment assignment for a single individual}. The model depicted in (a) has larger outcome prediction errors than the model depicted in (b), because the dashed lines in (a) are larger than the dashed lines in (b). However, the model in (a) makes a better treatment assignment than the model in (b), because the red dots preserve the ranking of the blue dots. }
    \label{fig:outcome}
\end{figure*}

We illustrate this with an example in Figure~\ref{fig:outcome}, which compares the outcome predictions made by two different models for a single individual. One model has high prediction errors (Figure \ref{fig:outcome-high}) and the other has low prediction errors (Figure \ref{fig:outcome-low}). The blue (dark) dots correspond to the true conditional expectations (they are the same for both figures), whereas the red dots correspond to the predictions. A larger distance between the blue and the red dots (represented by dashed lines) implies that the model makes worse outcome predictions. 

In this example, the conditional expectation when $T=1$ is larger than when $T=2$ (as shown by the blue dots), which implies that $T=1$ is a better treatment assignment. Therefore, models make the optimal assignment when $\hat{\mu}(x, 1)>\hat{\mu}(x, 2)$. Figure~\ref{fig:outcome-high} shows that the model with larger prediction errors actually makes the optimal treatment assignment because the rank ordering of the predicted outcomes across treatments is the same as the rank ordering of the true values. On the other hand, the second model depicted in Figure~\ref{fig:outcome-low} makes a worse assignment, even though its prediction errors are smaller, because the ordering is inverted. Importantly, this can also occur when fitting models for causal effect prediction. Therefore, since better outcome or causal effect prediction can lead to worse treatment assignments, learning models that predict optimal assignments (i.e., using an A-learner) should in principle lead to better assignments than the other two metalearners.

As a second main contribution, we empirically assess the practical impact that this analytical finding can have on treatment assignment performance by conducting a massive-scale, experimental comparison of the three metalearners in the context of content selection at Spotify. Our focal application is choosing, for each listener, which playlist generation algorithm (treatment) to apply in order to maximize the number of songs streamed. To our knowledge, this is the first real-world, at-scale comparison of the three metalearners. 

The experiment shows several things. First, it supports the analytical finding that models specifically trained to predict the best treatment are best for treatment assignment (i.e., the A-learner outperforms the O-learner and the E-learner for treatment assignment). This is the case even with training data consisting of more than half a billion observations, a surprising finding given that, in theory, all metalearners should converge to the same treatment assignment policy with large enough data. This finding also implies that we should reconsider some of the justifications made in prior research for using causal effect prediction methods for personalized treatment assignment. 

The experiment also reveals interesting findings for the application to content selection in music streaming.  Specifically, it shows that (1) using an algorithm assignment policy can substantially improve total streaming compared to the typical approach of applying the same playlist generation algorithm to everyone, and that (2) larger data sets lead to significantly better policies, illustrating the advantages of conducting massive-scale A/B tests for the purpose of learning treatment assignment policies (rather than just for evaluation or for selecting the best variant).

% and even more so when the predictions are made using a large number of features or models are estimated using algorithms that are prone to errors due to variance

\section{Treatment assignment problem}\label{sec:problem}

Treatment assignment problems correspond to settings where a decision-maker wants to maximize the overall causal effect of decisions on an outcome of interest (e.g., deciding what playlist generation algorithm to use for each listener to maximize the number of streams). Each possible alternative corresponds to a different treatment, and the goal is to assign individuals to the treatment that maximizes their outcome. We formalize the problem in this section. 

We consider settings in which decisions are independent and the treatment assignment policy is learned from historical data on previous decisions made at random. This implies that each decision affects a single unit (or instance) and there is no selection bias in the data. In the causal inference literature, the first assumption is also known as the \textbf{Stable Unit Treatment Value Assumption (SUTVA)}~\citep{cox1958planning}. The second assumption is known as \textbf{unconfoundedness} and implies there are no systematic differences between users assigned to different treatments. Unconfoundedness also goes by other names in different fields, including ignorability~\citep{rosenbaum1983central}, the back-door criterion~\citep{pearl2009causality}, and exogeneity~\citep{wooldridge2015introductory}.  Practically speaking, these assumptions hold when we use a carefully designed randomized A/B test to gather data.

Let $T$ be the treatment assignment variable and $Y$ be the observed outcome. We use potential outcomes to frame causality~\citep{rubin1974estimating} and define $Y(j)$ as the outcome we would observe if we were to assign treatment $j$ (out of $k$ possible treatment alternatives), so that $Y=Y(j)$ if $T=j$. Then, the treatment assignment that would lead to the best outcome on average is:
\begin{equation}
a^*=\argmax_j ~\mathbb{E}[Y(j)]
\end{equation}
and can be estimated using the sample mean ($\hat{\mathbb{E}}$) for each treatment: 
\begin{equation}\label{eq:sample_mean}
\hat{a}=\argmax_j ~\hat{\mathbb{E}}[Y|T=j]
\end{equation}

Equation~\ref{eq:sample_mean} describes a standard A/B test approach that compares multiple treatments across a predefined population. This approach, however, does not address individualized treatment assignments. In our setting, we \emph{learn} optimal treatments for individuals or subpopulations.  Suppose individuals vary with respect to a set of variables (features) $X$.  We can then think of a feature vector $x$ as a subpopulation where $X=x$ and formulate the optimal assignment (given $x$) as:
\begin{equation}\label{eq:optimal_policy}
a^*(x)=\argmax_j ~\mathbb{E}[Y(j)|X=x]
\end{equation}

Without the argmax, the right-hand side is essentially the formulation of a predictive model.  Applying statistical modeling frees us from specifying in advance what are the particular subpopulations of interest. In Section~\ref{sec:definition}, we present three metalearners that can be used to estimate $a^*(x)$ from data, each producing a treatment assignment policy $\hat{a}(x)$.

Treatment assignment policies can be evaluated in terms of their ability to minimize the expected difference between the outcome when optimal assignments are made, $Y(a^*(X))$, and the outcome when the policy is deployed, $Y(\hat{a}(X))$. This evaluation measure is also known as \textbf{expected regret} (or just `regret') in decision theory:
\begin{equation}\label{eq:min_measure}
\text{Regret}(\hat{a}) =\mathbb{E}_{Y(a^*(X)),Y(\hat{a}(X))}[Y(a^*(X))-Y(\hat{a}(X))],
\end{equation}
and minimizing regret is the same as maximizing the expected outcome of deploying the policy:
\begin{equation}\label{eq:max_measure}
\mathbb{E}_{Y(\hat{a}(X))}[Y(\hat{a}(X))]
\end{equation}

However, evaluating treatment assignment policies using historical data (as is typical when building standard machine learning models) is challenging because we do not observe all potential outcomes for any given individual; we only observe the single potential outcome for the treatment actually given. Therefore, if (for any given individual) the policy assigns a different treatment from the one that was assigned in the historical data, we do not know the corresponding potential outcome. Fortunately, given a data set of $n$ individuals from a randomized A/B test, we can still obtain an unbiased estimate of Equation~\ref{eq:max_measure}~\citep{li2010contextual}: 
\begin{equation}\label{eq:empirical_measure}
    \frac{1}{n}\sum_{i=1}^n \textbf{1}(\hat{a}(x_i)=t_i)\frac{y_i}{\mathbb{P}(T=t_i)},
\end{equation}
where for each individual $i$, $x_i$ is the feature vector, $t_i$ is the assigned treatment in the data, $y_i$ is the observed outcome, and $\mathbb{P}(T=t_i)$ is the probability of being assigned to treatment $t_i$ in the data (a known quantity if the data was collected through a randomized A/B test). We present a simplified proof of this result in Appendix~\ref{app:proof1} (see~\cite{li2010contextual} for a detailed proof).
 
\section{Metalearners}~\label{sec:definition}
 The causal inference literature often focuses its attention on the estimation of aggregate causal effects, such as the so-called \textbf{average treatment (or causal) effect (ATE)}, which corresponds to the average effect of a treatment across the individuals in some well-defined population. However, estimating the ATE does not help us to target different individuals with different treatments, because it does not discriminate between the individuals in the population at all. A fundamental motivation behind this work is that the population exhibits \textbf{heterogeneous treatment effects (HTEs)}, which are defined in terms of the degree to which a treatment may have different effects on different individuals~\citep{imai2013estimating}. 

 One can account for HTEs through the estimation of \textbf{conditional average treatment effects (CATEs)}, which correspond to the average causal effect conditioned on a set of available features. Thus, to the extent that individuals in the population differ on their features (and those features are related to causal effects), we may estimate different causal effects for each individual. Of course, treatment effects may still vary among individuals that share the same features (since we may not be accounting for all aspects related to the causal effect).  Nevertheless, the estimation of HTEs by using CATEs allows us to make different interventions for different individuals without knowing the relevant subpopulations in advance.
 
As we discuss in more detail in Section~\ref{sec:review}, the ideas behind CATE estimation have been fundamental to the development of methods for learning treatment assignment policies from data. We group such methods into three classes of algorithms (or metalearners), described below and in Table~\ref{tab:metalearners}. This grouping is meant to highlight the perspective that outcome prediction, causal effect prediction, and treatment assignment are different tasks, which has important implications for modeling and the use of predictive models for decision making. 
 
 \begin{enumerate}
     \item \textbf{Outcome Learner (O-learner):} Standard machine learning is used to learn a model that is optimized to predict the outcome under each treatment. The policy assigns individuals to the treatment with the largest predicted outcome.
    \item \textbf{Effect Learner (E-learner):} a machine learning method specifically designed to estimate CATEs is used to learn a model optimized to predict the \emph{differences} between the outcome of each treatment and a baseline (or control). The policy assigns individuals to the treatment with the largest predicted difference (i.e., treatment effect).
    \item \textbf{Assignment Learner (A-learner):} The problem of estimating optimal treatment assignments can be transformed into a weighted classification problem wherein each treatment corresponds to a class and the optimal classifier corresponds to the optimal treatment assignment policy~\citep{zadrozny2003policy,zhang2012estimating}. Weights are often defined using the observed outcome under each treatment condition, so that weights are larger for treatments with larger outcomes. Thus, standard machine learning can be used to learn a weighted classification model optimized to predict the treatment with the largest weight. The resulting classifier can then be used to assign individuals to the class (treatment) that is predicted to have the largest weight (outcome). 
\end{enumerate}

Section~\ref{sec:review} discusses multiple studies that have used or recommended specific instances of these metalearners for treatment assignment. Importantly, the three metalearners converge to optimal treatment assignments with large enough samples, assuming the machine learning procedure that is used to learn the models is a consistent estimator of the estimands presented in Table~\ref{tab:metalearners}. However, there are two key differences between the metalearners that are critical in practice, as summarized in Table~\ref{tab:analytical_comparison} and discussed next.

\begin{table}
    \renewcommand{\arraystretch}{0.7} 
    \centering
    \begin{tabular}{c|c|c|c|c}
        \toprule
        \multirow{2}{*}{\textbf{Metalearners}}& \multicolumn{3}{c|}{\textbf{Model may be used to predict:}} & \textbf{Learning procedure}\\
        \cline{2-4}
        & Outcomes & Effects & Assignments & \textbf{opmitized for} \\
        \midrule
         Outcome Learner & \multirow{2}{*}{\cmark}& \multirow{2}{*}{\cmark}& \multirow{2}{*}{\cmark} & Outcomes \vspace{-5pt}\\
         (O-learner) &&&& ($MSE_\mu$ in  Equation~\ref{eq:mse_mu})\\
         \hline
         Effect Learner & \multirow{2}{*}{\xmark}& \multirow{2}{*}{\cmark}& \multirow{2}{*}{\cmark}& Causal effects \vspace{-5pt}\\ 
         (E-learner)&&&& ($MSE_\tau$ in Equation~\ref{eq:mse_t})\\
         \hline
         Assignment Learner & \multirow{2}{*}{\xmark}& \multirow{2}{*}{\xmark}& \multirow{2}{*}{\cmark} & Assignments \vspace{-5pt}\\ 
         (A-learner)&&&& ($WMR$ in Equation~\ref{eq:wmr})\\
         \bottomrule
    \end{tabular}
    \caption{Comparison of metalearners for estimating treatment assignment policies}
    \label{tab:analytical_comparison}
\end{table}

\subsection{Distinction 1: Level of generality}\label{sec:generality}

The first key distinction between the three metalearners is their level of generality (the metalearners are listed above from the most general to the least general). O-learners are the most general of the metalearners because they produce models that predict outcomes, and such models may also be used to predict causal effects or optimal treatments. Specifically, causal effect predictions may be obtained by taking the difference between the predicted outcomes of two treatments under consideration, and optimal-treatment predictions may be obtained by selecting the treatment with the largest predicted outcome. Therefore, O-learners may be used for all three different purposes. 

Models that predict causal effects cannot be used to predict outcomes. For such models, predictions estimate the expected  marginal change in the outcome that results from assigning some specific treatment, but the predictions cannot be used to estimate expected outcomes under an arbitrary treatment condition. Therefore, while causal effect predictions may still be used to predict optimal treatments (by selecting the treatment with the largest predicted effect), E-learners are not as general as O-learners. 

Finally, models trained to predict optimal treatment assignments (i.e., learned with an A-learner) can only be used for that purpose. These models, the least general, cannot predict the outcome or the effect that would result from making those assignments.

This distinction implies that more general metalearners (O-learners and E-learners), can be preferable over A-learners when outcome and causal effect predictions are important for other reasons besides their usefulness to determine the treatment with the most beneficial outcome. For example,~\cite{mcfowland2021prescriptive} consider treatment assignment settings where there are budget constraints and the decision maker faces costs that are unknown ex-ante. In such settings, quantifying the benefit of each individual decision (e.g., via causal effect prediction) and the cost of each possible course of action (e.g., via outcome prediction) is important to allocate resources in the most profitable way. 

Feasible treatment assignment rules can also be constrained for ethical, legislative, or political reasons. For example, a public policy maker may want to prioritize the assignment of subsidies to individuals in some protected class unless the predicted effect of the subsidy on annual income is below a certain threshold or the individual is predicted to already have an annual income above certain threshold. Since assessing whether an individual meets these two conditions would require causal effect and outcome predictions, implementing an A-learner in this type of settings may be counterproductive or infeasible. 

A-learner decision rules may also be more difficult to implement in settings where the models are intended to support (rather than automate) human decision making. For example, for economic policy and medical treatment assignment, decision makers may need to weigh the potential benefit of the treatment alternatives with respect to some other information not available to the model (e.g., how the individuals affected by the treatments feel about the treatment alternatives), so predicting the treatment with the ``most beneficial outcome'' may not suffice. Additionally, information about causal effects and outcomes can be important for other reasons beyond decision making (e.g., for users to trust the model, to debug the model, to develop more effective treatments in the future).

%In other circumstances, A-learners may be problematic because the treatment may not correspond to something that can be assigned by the model. For instance, platforms like LinkedIn may be interested in offering job advice to their users to improve their career opportunities. In such an example, an A-learner could be used to identify the treatments that are (allegedly) the most beneficial, but the treatment assignments are ultimately up to the users, who are likely to be interested in the predicted outcome of each possible treatment alternative. Finally, in

Nonetheless, in settings where outcomes and effects are relevant only for the model to assign individuals to the most beneficial treatment, we should expect A-learners to make better treatment assignments because they are specifically designed for treatment assignment; we elaborate on this premise in detail in the rest of the paper. 

\subsection{Distinction 2: Learning procedure}

The second key distinction is that each metalearner uses a different learning objective (or loss function) for the machine learning. O-learners use a loss function designed to optimize outcome predictions; E-learners use a loss function designed to optimize causal effect predictions, and A-learners use a loss function designed to optimize treatment assignments. This implies that, while all metalearners share the same ultimate goal (optimizing treatment assignments as specified by Equations~\ref{eq:min_measure},~\ref{eq:max_measure}, and~\ref{eq:empirical_measure}), they differ with respect to the procedures they use to learn from data.

This distinction is important because an improvement in the prediction of outcomes or causal effects does not imply an improvement in treatment assignment (as previously shown in Figure~\ref{fig:outcome}). In fact, the improvements may occur at the expense of worse treatment assignments! Thus, we should expect machine learning with loss functions specifically tailored to optimize treatment assignments to produce better models: the A-learner should outperform the other metalearners with finite training data when making treatment assignment decisions.

Nevertheless, O-learners and E-learners are much more commonly used among scholars and practitioners in marketing and information systems (IS), even though those metalearners are optimized to minimize prediction errors in outcomes or causal effects rather than decision making errors. One goal of this study is to encourage a more widespread consideration, study, and use of A-learners among management researchers by showing how decisions can be substantially improved when machine learning models are directly optimized for treatment assignment (decision making).

\section{Choice of objective function}

In this section, we compare the three metalearners analytically to illustrate how their choice of objective function may affect their performance in treatment assignments.

\subsection{Outcome prediction}\label{sec:ol_desc}

As mentioned, the O-learner assigns treatments by learning one or more models that predict the expected outcome of each treatment ($\hat{\mu}$):
\begin{equation}\label{eq:outcome}
\hat{\mu}(x, j) = \hat{\mathbb{E}}[Y|X=x,T=j],
\end{equation}
and then selecting the treatment with the best predicted outcome:\footnote{Note that this policy does not use confidence intervals to make treatment assignment decisions; only the point estimates are used. This is because, from the perspective of a regret minimizer, the best possible choice (in expectation) is the treatment with the largest point estimate, regardless of the confidence intervals. For this reason, confidence intervals are not incorporated as part of the decision making in our study.}
\begin{equation}\label{eq:choose_outcome}
\hat{a}_\mu(x)=\argmax_j ~\hat{\mu}(x, j)
\end{equation}

A standard approach to fit Equation~\ref{eq:outcome} is to regress outcome $Y$ on features $X$ and treatment assignment $T$ using various machine learning methods designed to minimize the mean squared error for the outcome ($MSE_\mu$):
\begin{equation}\label{eq:mse_mu}
MSE_\mu(\hat{\mu},j) = \mathbb{E}_{X,Y}[(Y-\hat{\mu}(X, j))^2|T=j],
\end{equation}
and then to choose the model(s) with the lowest $MSE_\mu$.

The premise here is that minimizing $MSE_\mu$ implies better outcome predictions, and therefore better treatment assignments. However, optimizing for outcome prediction (by minimizing $MSE_\mu$ or other measures such as mean absolute error or cross-entropy) does not necessarily optimize for treatment assignment. Going back to the earlier example in the introduction, Figure~\ref{fig:outcome-high} shows that the model with larger prediction errors makes the optimal treatment assignment because the rank ordering of the predicted outcomes is the same as the rank ordering of the true values. The second model makes a worse assignment, even though its prediction errors are smaller, because the ordering is inverted. Therefore, choosing the model with the lower (and thus better) $MSE_\mu$ leads to a worse treatment assignment.

Multiple researchers have noted the potential of overfitting when multiple outcome models are used to estimate treatment effects instead of directly fitting a causal effect model~\citep{kunzel2019metalearners,nie2017quasi}. For example, suppose that features $X_1$ and $X_2$ are predictive of outcomes, but only feature $X_1$ is predictive of effects. This implies that, for the purposes of estimating effects and assigning treatments, segmenting individuals using exclusively $X_1$ is more statistically efficient than segmenting them according to $X_1$ and $X_2$. Hence, by focusing statistical power on features that are predictive of effects, models optimized for treatment effect estimation can achieve lower bias and lower variance than models optimized for outcome prediction. Subsequently, many researchers have proposed methods that directly model effects (rather than outcomes) to make treatment assignments; these are instances of the E-learner. However, as we discuss next, optimizing for causal effects is not the same as optimizing for treatment assignments either.

\subsection{Causal effect prediction}\label{sec:cp_desc}

The second metalearner, the E-learner, consists of learning one or more models to estimate the CATEs ($\hat{\tau}$) directly:
\begin{equation}\label{eq:effect}
\hat{\tau}(x, j) = \hat{\mathbb{E}}[Y|X=x,T=j] - \hat{\mathbb{E}}[Y|X=x,T=0],
\end{equation}
where $T=0$ corresponds to a baseline treatment (e.g., the control in an A/B test setting). The optimal treatment may then be chosen as follows:
\begin{equation}\label{eq:choose_causal}
\hat{a}_\tau(x)=\argmax_j ~\hat{\tau}(x, j)
\end{equation}

The (sometimes unstated) goal of machine learning methods designed for the estimation of CATEs is to minimize the mean squared error for treatment effects ($MSE_\tau$):
\begin{equation}\label{eq:mse_t}
MSE_\tau(\hat{\tau}, j) = \mathbb{E}_{X,Y(j),Y(0)}[(Y(j)-Y(0)-\hat{\tau}(X, j))^2] 
\end{equation}

Therefore, these methods are not optimized to predict outcomes but rather to predict causal effects, which are usually defined as the difference between potential outcomes (i.e., $Y(1)-Y(0)$). The main challenge is that we only observe one potential outcome for any given individual, so we cannot calculate Equation~\ref{eq:mse_t} directly. However, we may use alternative formulations to estimate $MSE_\tau$ from data~\citep{schuler2018comparison}, allowing us to compare (and optimize) models on the basis of how good they are at predicting causal effects. 

\begin{figure*}
    \centering
    \begin{subfigure}{.49\textwidth}
    \centering
     \begin{tikzpicture}[scale=0.8]
        \linespread{0.5}
        \begin{axis}[
            %hide y axis,
            ylabel = {Causal effect},
            ymax=4,
            ymin=-1,
            xmax= 2.5,
            xmin = 0.5,
            xticklabels={$T=1$ (better), $T=2$ (worse)}, xtick={1,2},
            axis y line*=left,
            axis x line*=bottom]
        \addplot[only marks, mark=*,red] plot coordinates {
            (1, 3.7)
            (2, -0.7)};
        \addplot[only marks, mark=*,blue] plot coordinates {
            (1, 2)
            (2, 1)};
        \node[align=center] at (axis cs:1,0.2) {Assigned \\ Treatment};
        \addplot[style={thick},->] coordinates {(1,-0.3) (1,-0.7)};
        \addplot[no marks,dashed] plot coordinates {
            (1, 3.7)
            (1, 2)};
        \addplot[no marks,dashed] plot coordinates {
            (2, 1)
            (2, -0.7)};
    \end{axis}
    \end{tikzpicture}
    \caption{Model with worse $\hat{\tau}$}
    \label{fig:effect-high}
    \end{subfigure}
    \begin{subfigure}{.49\textwidth}
    \centering
     \begin{tikzpicture}[scale=0.8]
        \linespread{0.5}
        \begin{axis}[
            %hide y axis,
            ylabel = {Causal effect},
            ymax=4,
            ymin=-1,
            xmax= 2.5,
            xmin = 0.5,
            xticklabels={$T=1$ (better), $T=2$ (worse)}, xtick={1,2},
            axis y line*=left,
            axis x line*=bottom]
        \addplot[only marks, mark=*,red] plot coordinates {
            (1, 1.2)
            (2, 1.8)};
        \addlegendentry{~Predictions ($\hat{\tau}$)}
        \addplot[only marks, mark=*,blue] plot coordinates {
            (1, 2)
            (2, 1)};
        \addlegendentry{~True values ($\tau$)}
        \addplot[no marks,dashed] plot coordinates {
            (2, 1.8)
            (2, 1)};
        \addplot[no marks,dashed] plot coordinates {
            (1, 2)
            (1, 1.2)};
        \addlegendentry{~Prediction error}
        \node[align=center] at (axis cs:2,0.2) {Assigned \\ Treatment};
        \addplot[style={thick},->] coordinates {(2,-0.3) (2,-0.7)};
    \end{axis}
    \end{tikzpicture}
    \caption{Model with better $\hat{\tau}$}
    \label{fig:effect-low}
    \end{subfigure}
    \caption{\textbf{Comparison of causal effect prediction vs treatment assignment for a single individual}. The model depicted in (a) has larger effect prediction errors than the model depicted in (b), because the dashed lines in (a) are larger than the dashed lines in (b). However, the model in (a) makes a better treatment assignment than the model in (b), because the red dots preserve the ranking of the blue dots. }
    \label{fig:effect}
\end{figure*}

Unfortunately for our application, and similarly to the previous section, optimizing causal effect predictions (by minimizing $MSE_\tau$) is not the same as optimizing treatment assignments either. We illustrate this in Figure~\ref{fig:effect}, which shows a similar example to the one illustrated in Figure~\ref{fig:outcome}, except that it compares the causal effect (rather than outcome) predictions made by two models.\footnote{In this case, the treatments $T=1$ and $T=2$ are being compared with a baseline treatment $T=0$. Therefore, the y-axis in Figure~\ref{fig:effect} represents the difference in outcomes with respect to $T=0$.} Therefore, the blue (dark) dots in this example represent the causal effects of the treatments for a specific individual (these dots are the same in both graphs), and the red dots represent the estimation of the effects by the models. As before, the first model has high prediction errors (Figure \ref{fig:effect-high}) but makes a better assignment, while the second has lower prediction errors (Figure \ref{fig:effect-low}) but makes a worse assignment. Thus, the model that makes a better causal effect prediction (i.e., that has lower $MSE_\tau$) makes a worse treatment assignment. 

Surprisingly, this implies that models that are (relatively) bad at causal effect prediction may be good at making treatment assignments. This result, while seemingly counter-intuitive at first, may be attributed to the bias-variance decomposition of errors. In the machine learning community, it is well known that models that have a good classification performance are not necessarily good at estimating class probabilities, and vice versa~\citep{friedman1997bias}. A useful analogy in our context is to think about treatment assignment as a classification problem and to think about causal effect estimation as a probability estimation problem; the two tasks are closely related but not exactly the same. Importantly, the bias and variance components of the estimation error in causal effect predictions may combine to influence treatment assignment errors in a very
different way than with the squared error of the predictions themselves~\citep{fernandez2019causal}. 

\pgfmathdeclarefunction{gauss}{2}{%
  \pgfmathparse{1/(#2*sqrt(2*pi))*exp(-((x-#1)^2)/(2*#2^2))}%
}

\begin{figure}
\centering
\begin{subfigure}{.49\textwidth}
  \centering
  \begin{tikzpicture}[scale=0.7]
    \begin{axis}[every axis plot post/.append style={
      mark=none,domain=-2.5:5.5,samples=50,smooth},
        % All plots: from -2:4, 50 samples, smooth, no marks
      axis x line*=bottom, % no box around the plot, only x and y axis
      xtick={-2, -1, 0, 1, 2, 3, 4, 5},
      xlabel ={Causal effect},
      xticklabel style={font=\small},
      xticklabels={-2, -1, 0, $\tau(2)$, $\tau(1)$, 3, 4, 5},
      hide y axis, ymax=1.2,
      tick label style={font=\Large},
      axis y line*=left, % the * suppresses the arrow tips
      enlargelimits=false] % extend the axes a bit to the right and top
      \addplot {gauss(-0.7,0.5)};
      \addlegendentry{Sampling distribution of $\hat{\tau}(x, 2)$}
      \addplot {gauss(3.7,0.5)};
      \addlegendentry{Sampling distribution of $\hat{\tau}(x, 1)$}
      %\addlegendentry{OM}
   \end{axis}
    \end{tikzpicture}
  \caption{Model with worse $\hat{\tau}$}
  \label{fig:bias_high}
\end{subfigure}%
\begin{subfigure}{.49\textwidth}
  \centering
  \begin{tikzpicture}[scale=0.7]
  \begin{axis}[every axis plot post/.append style={
      mark=none,domain=-2.5:5.5,samples=50,smooth},
        % All plots: from -2:4, 50 samples, smooth, no marks
      axis x line*=bottom, % no box around the plot, only x and y axis
      xtick={-2, -1, 0, 1, 2, 3, 4, 5},
      xlabel ={Causal effect},
      xticklabel style={font=\small},
      xticklabels={-2, -1, 0, $\tau(2)$, $\tau(1)$, 3, 4, 5},
      hide y axis, ymax=1.2,
      tick label style={font=\Large},
      axis y line*=left, % the * suppresses the arrow tips
      enlargelimits=false] % extend the axes a bit to the right and top
      %\addlegendentry{OM}
      \addplot {gauss(1,0.5)};
      \addlegendentry{Sampling distribution of $\hat{\tau}(x, 2)$}
      %\addlegendentry{EM}
      \addplot {gauss(2,0.5)};
      \addlegendentry{Sampling distribution of $\hat{\tau}(x, 1)$}
   \end{axis}
   \end{tikzpicture}
  \caption{Model with better $\hat{\tau}$}
  \label{fig:bias_small}
\end{subfigure}
\vspace{3mm}
\caption{Sampling distributions of \boldmath$\hat{\tau}$. The model depicted in (a) is a biased estimator of causal effects, whereas the model depicted in (b) is unbiased. However, the model in (b) is much more likely to estimate that $\hat{\tau}(x,1)<\hat{\tau}(x,2)$ (and thus make the wrong assignment) because of sampling error.}
\label{fig:bias}
\end{figure}

Figure~\ref{fig:bias} illustrates this in more detail by depicting the sampling distribution of the causal effect estimates previously shown in Figure~\ref{fig:effect}. Specifically, Figure~\ref{fig:bias_high} shows that the large prediction errors of the model in Figure~\ref{fig:effect-high} are the result of high bias because the sampling distributions are not centered on the causal effect estimands. However, this model works very well for treatment assignment because  $\mathbb{P}(\hat{\tau}(x,2)<\hat{\tau}(x,1))\approx1$ and $\tau(x,2)<\tau(x,1)$. On the other hand, Figure~\ref{fig:bias_small} shows that the model in Figure~\ref{fig:effect-low} is an unbiased estimator of causal effects and has lower mean squared error. However, this model is more likely to make the incorrect assignment due to sampling errors (variance). Importantly, what matters in this case is not the accuracy of the causal effect estimates but how good they are at discriminating between treatment alternatives. We elaborate more on this in Section~\ref{sec:obj_discussion}, after discussing the objective function of the A-learner.

\subsection{Treatment assignment prediction}\label{sec:ap_desc}

The third metalearner, the A-learner, estimates the treatment assignment policy by directly learning the treatment assignments that lead to the best outcomes. As \cite{zhang2012estimating} describe in detail, the treatment assignment problem can be transformed into a weighted classification problem. The idea is that each treatment alternative can be mapped to a class, and classes are associated with weights that correspond to the cost of not predicting the corresponding class. Weights are generally defined in terms of the potential outcomes associated with each treatment alternative~\citep{beygelzimer2009offset,zhao2012estimating,kitagawa2018should}, but more broad definitions exist~\citep{zhang2012estimating}. Thus, the goal is to `classify' individuals into the class (treatment) with the largest weight in order to minimize misclassification costs.

The challenge here is that we only observe a single weight for any given individual (the one associated with the treatment assigned in the data), so we do not observe correct classifications at the individual level. Fortunately, samples from treatment assignment problems can be transformed into weighted classification samples so that any importance-weighted classification algorithm can be used to learn treatment assignment policies. For example, given a probability distribution $\mathbb{P}(T)$ over the treatment (e.g., the probability that an individual gets assigned to treatment $T$ in the A/B test data), each observation ($x, y, t$) can be transformed into an importance-weighted example where $y/\mathbb{P}(t)$ is the cost of not predicting treatment $t$ given input $x$~\citep{beygelzimer2009offset,zhao2012estimating}; if $t$ is predicted, then the cost is zero. The weighted misclassification rate ($WMR$) of classifier $\hat{a}$ under this setting is:
\begin{equation}\label{eq:wmr}
        WMR(\hat{a}) = \mathbb{E}_{X,Y,T}\Bigg[\textbf{1}(\hat{a}(X)\neq T)\frac{Y}{\mathbb{P}(T)}\Bigg],
\end{equation}
which is directly tied to treatment assignment performance because minimizing the $WMR$ is equivalent to minimizing expected regret (as defined in Equation~\ref{eq:min_measure}). We present a simplified proof of this result in Appendix~\ref{app:proof2} (see~\cite{beygelzimer2009offset} for more details). As a result, we should expect $WMR$ to be a better objective function than $MSE_\mu$ or $MSE_\tau$ when the goal is to make the best possible treatment assignments. 

\subsection{Bias-variance tradeoff}\label{sec:obj_discussion}

\begin{figure}
    \centering
    \includegraphics[width=0.7\textwidth]{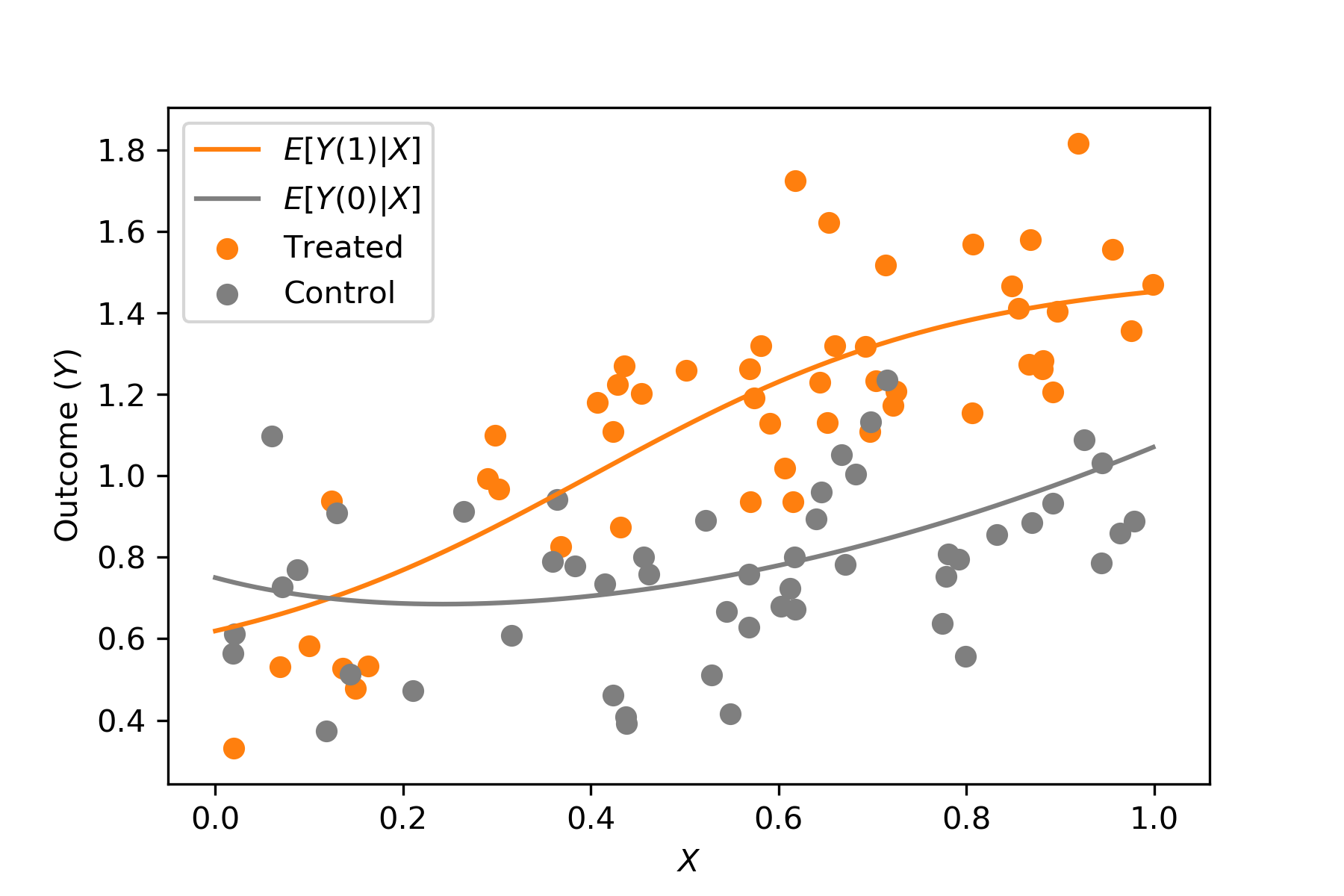}
    \caption{Sample data for treated and control individuals. The lines represent the expected potential outcomes for treated (orange) and untreated (gray), and the dots represent the data points in the sample. Individuals should be treated when the orange line is above the gray line and should not be treated otherwise.}
    \label{fig:ap_example}
\end{figure}

In this section, we use a simulated example to illustrate how the A-learner can exploit the bias-variance tradeoff in the learning procedure to make better treatment assignments than the E-learner. The data from the simulated example are shown in Figure~\ref{fig:ap_example}. There are two treatment alternatives, treat (orange) and not treat (gray), and the goal is to use feature $X$ to learn a treatment assignment policy to discriminate individuals into those that would benefit from the treatment (orange line is above gray line) and those that would not (gray line is above orange line). The lines represent the expected outcomes under both treatment conditions, and the dots represent the available training data. See Appendix~\ref{app:data} for a description of the data generating process.

Similarly to previous studies that have considered models with constrained functional forms for ethical, legislative, or political reasons~\citep{kitagawa2018should,athey2021policy}, suppose we are considering the class of treatment rules that split the population only once (e.g., decision trees with a single split). One alternative is to split individuals according to causal effect heterogeneity and then make treatment assignments according to their estimated causal effects, which corresponds to the E-learner. 

\begin{figure}
    \centering
    \includegraphics[width=0.7\textwidth]{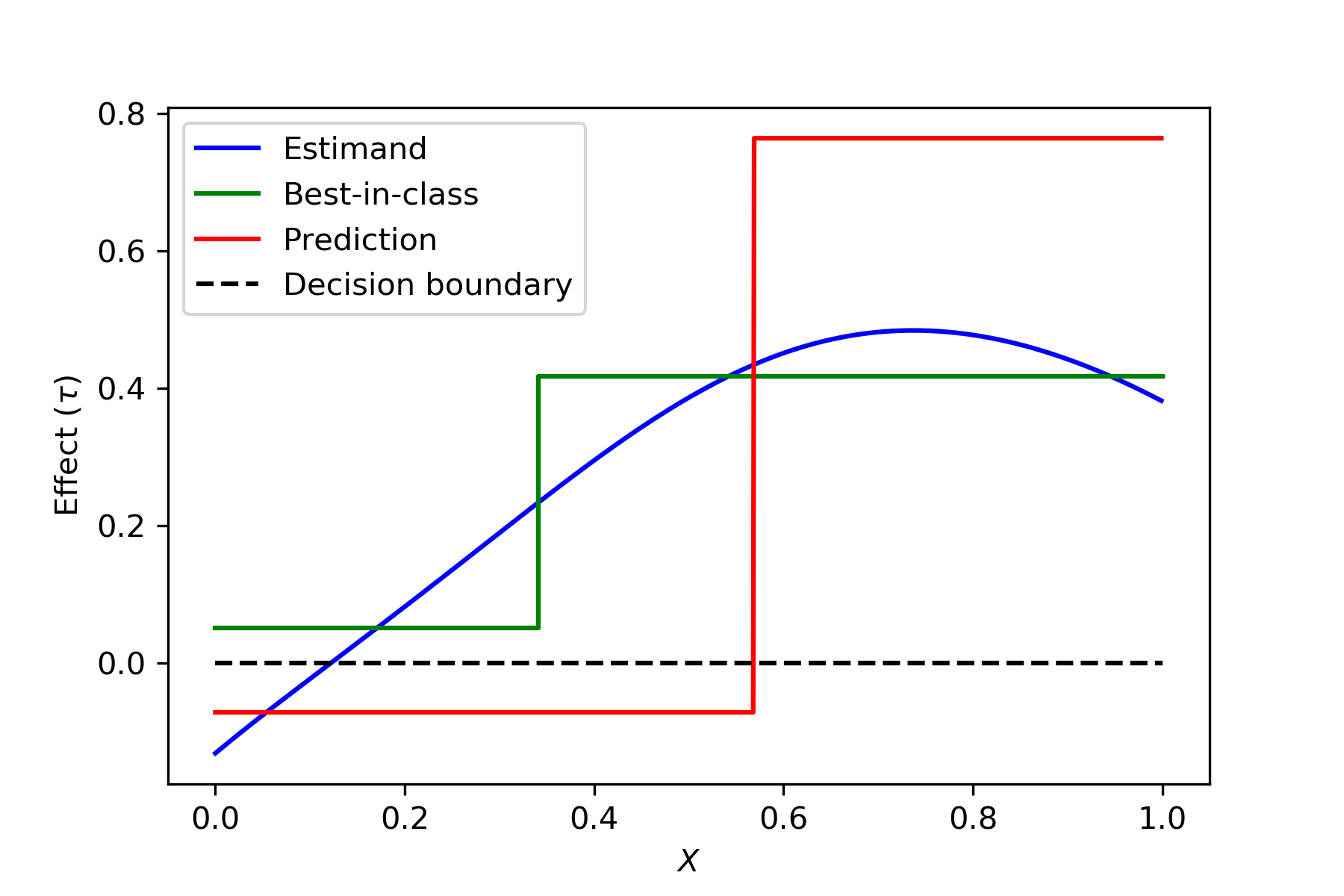}
    \caption{Estimation of causal effects vs. treatment assignment. The blue line is the expected causal effect, the red line is the estimated causal effect according to a single-split tree learned from data, and the green line is the best effect estimation that can be obtained using a single-split tree. Neither the estimated tree nor the best-in-class tree lead to optimal assignments.}
    \label{fig:cp_metalearner}
\end{figure}

Figure~\ref{fig:cp_metalearner} shows the result of learning a single-split causal tree---a tree that splits individuals according to causal effects~\citep{athey2016recursive}. The blue line is the expected treatment effect given $X$ (the estimand in causal effect estimation); the red line is the prediction from the tree that is learned from the data shown in Figure~\ref{fig:ap_example}; the green line is from the tree that is learned with unlimited data (i.e., the best possible single split that could be made to minimize $MSE_\tau$). Finally, the dashed line corresponds to the actual decision boundary: it is optimal to treat when the expected treatment effect (blue line) is above the boundary. The treatment assignment policy treats when the predicted effect (green or red lines) is above the boundary. 

Note that a treatment assignment policy based on estimated causal effects (red line) would not treat individuals with $X\lesssim0.6$ even though most individuals with such values for $X$ would actually benefit from the treatment. This is the result of variance in the estimation procedure. With more data, errors due to variance eventually disappear, and the estimated causal effects (red line) converge to the best-in-class predictor (green line). 

Unfortunately, the best-in-class predictor does not lead to optimal assignments either, because it estimates that the the treatment is beneficial for everyone even though individuals with a small value for $X$ do not benefit from the treatment. In this case, errors in treatment assignment occur due to bias in the estimation procedure: the causal tree is not complex enough to identify who does not benefit from the treatment. Nevertheless, this does not imply that the class of treatment rules that split the population only once is not complex enough to model treatment assignments. It just implies that learning a tree by splitting the population according to effect heterogeneity (an E-learner) does not lead to optimal assignments.

A second alternative is to use an A-learner to split the population according to \textit{preferred treatment assignments}. In other words, to learn a classification model optimized to minimize Equation~\ref{eq:wmr} ($WMR$) instead of Equation~\ref{eq:mse_t} ($MSE_\tau$). An important challenge, however, is that Equation~\ref{eq:wmr} can
be viewed as a weighted version of 0-1 loss, and it is well known in the machine learning literature that minimizing such loss is difficult due to its discontinuity and non-convexity.  A common approach to address this challenge is to use a surrogate loss to learn a scoring model, such as the negative log likelihood in logistic regression or the hinge loss in support vector machines~\citep{zhao2012estimating}, and then classify individuals according to their scores. The predictions of the resulting scoring model ($\hat{\theta}$) would correspond to:
\begin{equation}\label{eq:assignment}
\hat{\theta}(x, j) = \hat{\mathbb{P}}(\tilde{T}=j|X=x),
\end{equation}
where $\tilde{T}$ corresponds to the weighted treatment class, and the scoring model may be used to choose the optimal treatment as follows:
\begin{equation}\label{eq:choose_treatment}
\hat{a}_\theta(x)=\argmax_j ~\hat{\theta}(x, j)
\end{equation}

\begin{figure}
    \centering
    \includegraphics[width=0.7\textwidth]{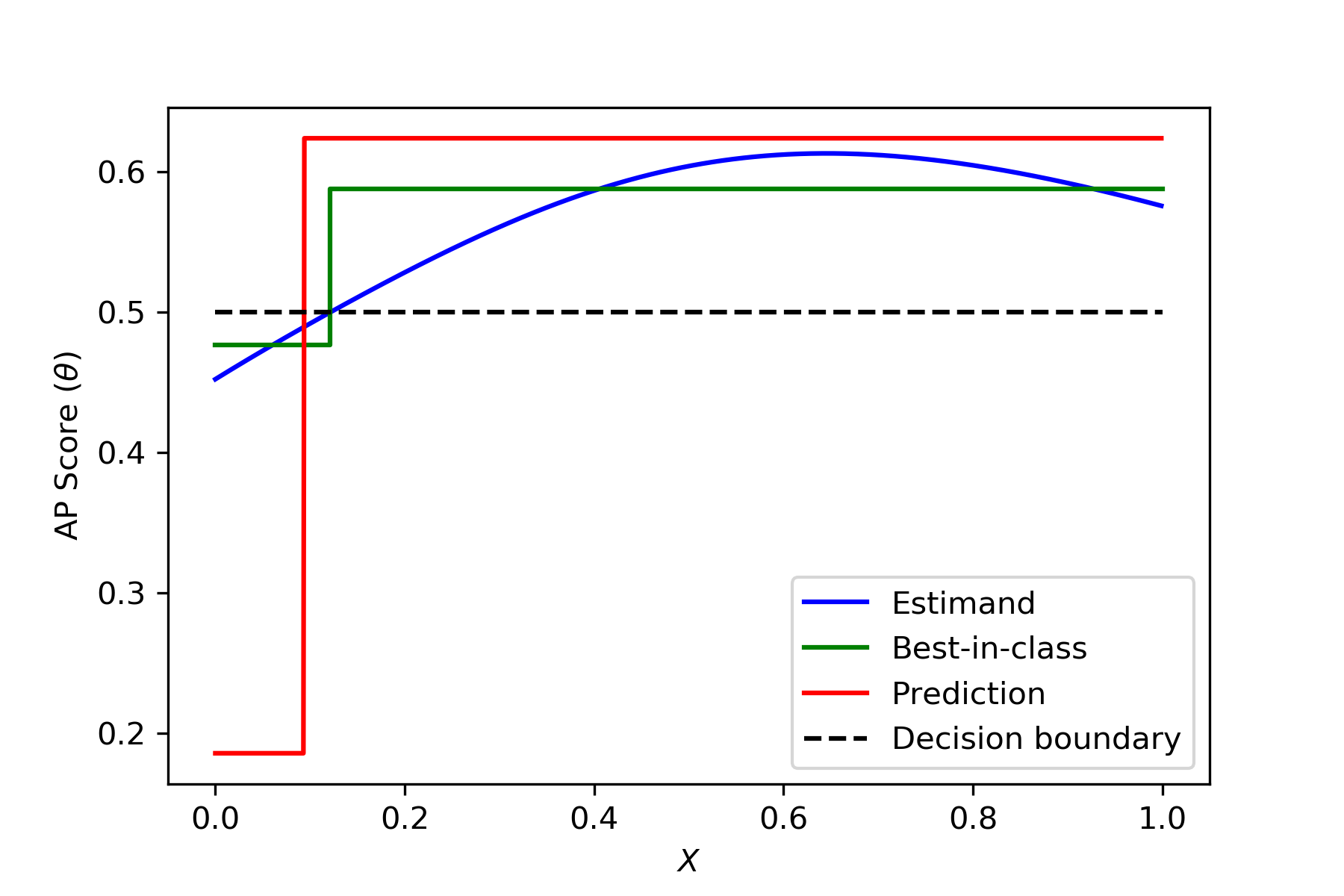}
    \caption{Learning trees designed to optimize for treatment assignment. The blue line is the target treatment score, the red line is the estimated score according to a single-split tree learned from data, and the green line is the best score estimation that can be obtained using a single-split tree. The estimated tree does a good job assigning treatments, and the best-in-class tree leads to optimal assignments even though it does not provide entirely accurate estimations of scores.}
    \label{fig:ap_metalearner}
\end{figure}

Nevertheless, in the case of tree-based algorithms, it is tractable to optimize according to 0-1 loss (and hence according to $WMR$). Figure~\ref{fig:ap_metalearner} shows the result of learning such a tree. Similarly to before, the blue line is the target score in the weighted classification task (the estimand); the red line is the scoring model that is learned from the data shown in Figure~\ref{fig:ap_example}; the green line is the scoring model that is learned with unlimited data (i.e., the best possible single split that could be made to minimize $WMR$); the dashed line corresponds to the decision boundary.

Note that although the estimated scoring model (red line) also suffers from errors due to variance and bias, these errors do not affect decision making (treatment assignment) as much. For example, there is a substantial underestimation of the treatment score for individuals with a small value for $X$, but this does not affect decision making because the optimal decision for those individuals is not to intervene. Importantly, with more data, the model eventually converges to the best-in-class predictor (green line), which leads to optimal assignments even though the treatment scores are biased due to the simplicity of the treatment rules under consideration.

Essentially, this example shows that minimizing errors in causal effect (or outcome) predictions may not imply better treatment assignments because the direction of the errors is critical. For the purposes of decision making, overestimations do not hurt when the treatment is beneficial, and underestimations do not hurt when the treatment is detrimental. This point is important because E-learners may correct such errors at the expense of increasing errors that will hurt decision making (as shown in Figure~\ref{fig:bias}). In contrast, A-learners are specifically designed to minimize errors that hurt decision making; errors that do not affect decisions are essentially ignored. 

In our example, the E-learner splits the data to minimize the bias in the causal effect predictions, whereas the A-learner splits the data to minimize the bias that negatively affects decisions. Table~\ref{tab:example} compares the two metalearners with unlimited training data (resulting in the green lines in Figures~\ref{fig:cp_metalearner} and \ref{fig:ap_metalearner}). The A-learner split leads to larger bias (and hence larger $MSE_\tau$) than the E-learner split, but that bias does not have negative implications for decision making (regret).

\begin{table}[]
    \renewcommand{\arraystretch}{0.7} 
    \centering
    \begin{tabular}{c|c|c|c}
    \toprule
    \textbf{Splitting criterion}&$\text{Squared bias}^*$&$MSE_\tau^\dagger$& Regret$^\dagger$ \\
    \midrule
         $MSE_\tau$ (E-learner split)  &0.007  &0.007    & 0.008   \\
         $WMR$ (A-learner split)       &0.019  &0.019    & 0       \\
        \bottomrule
        \multicolumn{4}{l}{\footnotesize * $\text{Squared bias}=\mathbb{E}[(\tau(X)-\mathbb{E}[\hat{\tau}(X)])^2]$}\\
        \multicolumn{4}{l}{\footnotesize $\dagger$ These measures exclude idiosyncratic noise and assume unlimited} \vspace{-5pt}\\
        \multicolumn{4}{l}{\footnotesize  data, so they are exclusively driven by bias.} \\
        \hline
    \end{tabular}
    \caption{Comparison of splitting criteria}
    \label{tab:example}
\end{table}

Of course, if decisions are based on \emph{true} causal effects rather than causal effect estimates (i.e., the blue line in Figure~\ref{fig:cp_metalearner}), then regret is also minimized. So, with more data, one could learn a more complex causal effect model (e.g., a tree with more splits), decrease the modeling bias, and eventually converge to optimal decision making. Nonetheless, as our subsequent empirical analysis shows, the A-learner can outperform the O-learner and the E-learner even when the training sample consists of hundreds of millions of observations.

\section{Experiments \& Results}

We now present an empirical comparison of the three metalearners for choosing which playlist generation algorithm to apply for each listener~\citep[see][for an overview of prior playlist generation studies in IS]{liebman2019right}.

\subsection{Application setting}

%A/B tests are experiments where two or more variants of a `treatment' are shown to users at random, and statistical analysis is used to determine which variant performs better for a given goal, which will will generically call "conversion" (more on this presently).  

In our playlist generation setting, the treatment variants consist of different algorithmic playlist generation systems tested in production by Spotify, a media services provider. Each system uses a different algorithm to select and rank songs in `algorithmic' playlists (playlists that are built dynamically according to user data).
The company has multiple goals when deploying such systems (e.g., converting users from free to premium, reducing churn, increasing engagement with the platform). However, the complexities of data collection, modeling, and deployment have historically made it prohibitively difficult for systems to be directly optimized in terms of these goals. Therefore, the models that underlie these systems are often heuristic (e.g., songs are ranked based on their similarity to other songs the user has played).

We focus specifically on the number of streamed songs in the playlist as a proxy for engagement and use it as our target outcome metric; this measure is significantly less noisy than other alternative engagement metrics and is available for all users.\footnote{Alternative engagement metrics, such as whether the user added the playlist to favorites, consist of actions that may not be available to all users and are generally harder to optimize due to their rare occurrence.} Thus, the goal is to assign users to the playlist generation system with which they would listen to the most songs. Firms typically run A/B tests to compare new machine-learned systems with the existing production system (as a baseline) and decide whether to replace the production system with one of the new systems. This essentially chooses the same treatment assignment for all users. However, as we have argued throughout this paper, different variants may work better for different users: if System A is best for new users and System B is best for more experienced users, then deploying the same system for all users would lead to sub-optimal treatment assignments. Since the outcome of interest in this case is song streams, one could then learn a treatment assignment policy that deploys different systems for different users in order to maximize the number of song streams.

In Appendix~\ref{app:diffs}, we provide a detailed discussion on how the deployment of content selection systems differs from other treatment assignment problems.

\subsection{Data}

We compare the three metalearners using data from a massive, production A/B test. The A/B test produced a data set in which four different playlist generation systems were randomly assigned to users to build algorithmic playlists: three newly developed playlist generation systems and the system that was currently in production. More specifically, each observation corresponds to a user who selected an algorithmic playlist, and each playlist was built using one of the four systems (chosen at random) to select and rank songs. 

There are 770 million observations in the data: 86.68\% assigned to the production system, and 4.44\% for each of the new variants. For each observation, we have the following categorical features: country (19 values), playlist ID (this serves as an identifier of the pool of songs that can be used to build the algorithmic playlist; 6 values), platform (e.g., Android; 3 values), user tenure in days (transformed into a discrete variable with 4 values), and product (e.g., free, premium; 8 values). For each categorical variable, the categories with fewer than 10,000,000 observations were grouped together in a category named `Other', resulting in the number of values for each variable reported above. Descriptive statistics for the features are shown in Table~\ref{tab:descriptive_stats}, and balance tests with respect to these features confirmed an adequate randomization of the systems. For each observation, we also have the number of total streams for the user (which is the target outcome).

Importantly, these massive data allow us to assess what would happen with data sets of many different sizes. Given a universal approximator (e.g., a tree-induction algorithm), all metalearners converge to the same (optimal) treatment assignment policy when the training data is large enough. So, the interest here is not to compare the metalearners when there is ``unlimited'' (i.e., very large) training data. Instead, the interest is to compare them across data sets of different sizes. 

\begin{table}[]
    \renewcommand{\arraystretch}{0.7} 
    \centering
    \begin{tabular}{c|c|c|c}
    \toprule
    \textbf{Categorical feature}&\textbf{\# of values}&\textbf{Entropy$^*$}&\textbf{Mode}\\
    \midrule
        Platform    &   3   &   0.471   &   iOS                 \\
        Country     &   19  &   0.596   &   United States       \\
        Playlist ID &   6   &   0.169   &   Other               \\
        Product     &   8   &   0.559   &   Free Subscription   \\
        User Tenure &   4   &   0.309   &   $>$179 days        \\
        
        \bottomrule
        \multicolumn{4}{l}{\footnotesize{$*$ Entropy was normalized to range between 0 and 1. Larger values imply a more}\vspace{-5pt}} \\
        \multicolumn{4}{l}{\footnotesize{uniform distribution of observations among feature values.}}\\
        \hline
    \end{tabular}
    \caption{Descriptive statistics of the categorical features available for treatment assignment}
    \label{tab:descriptive_stats}
\end{table}

\subsection{Learning and Evaluating Policies}\label{sec:learning_eval}

We compare treatment assignment policies estimated with each metalearner: (1) the \textbf{O-learner policy}, which assigns treatments based on a model that predicts the total number of streams for each system (Equation~\ref{eq:choose_outcome}); (2) the \textbf{E-learner policy}, which assigns treatments based on a model that predicts the increase in the total number of streams (compared to control) for each system (Equation~\ref{eq:choose_causal}); and (3) the \textbf{A-learner policy}, which uses a classification model to predict and assign the system that is estimated to produce the largest number of streams (Equation~\ref{eq:choose_treatment}). 

We use tree-based algorithms to learn all models so that differences in performance can be attributed to the loss functions used by each metalearner rather than the machine learning algorithm being used. We chose trees over other alternatives for multiple reasons. First, this choice allows us to demonstrate that the simulated example in Section~\ref{sec:obj_discussion} is not merely hypothetical: A-learner trees can indeed lead to substantially better assignments than E-learner trees. Second, tree models can be adjusted to predict outcomes, effects, and optimal assignments regardless of the metalearner; this will be important to compare the metalearners in Section~\ref{sec:imp_obj}. Finally, trees were the best performing models out of the multiple alternatives we considered. Appendix~\ref{app:extended} provides an extended analysis that also considers random forests and linear models. 

%Finally, we expect the difference in performance between metalearners to be more prominent with high-variance algorithms (such as tree induction algorithms) because in those cases it takes a larger training sample for the metalearners to converge. However, Section~\ref{sec:algorithm} also considers random forests and linear models as alternatives with lower variance.

The O-learner uses a decision tree regressor that minimizes $MSE_\mu$ (Equation~\ref{eq:mse_mu}) to learn $\hat{\mu}$ (Equation~\ref{eq:outcome}). The E-learner uses a decision tree regressor on the transformed variable proposed by~\cite{athey2016recursive} to learn $\hat{\tau}$ (Equation~\ref{eq:effect})  by minimizing $MSE_\tau$ (Equation~\ref{eq:mse_t}), i.e., a ``causal tree''. The E-learner policy uses 3 causal trees, one for each system except control (the regressor does not support non-binary treatments). Finally, the A-learner uses a weighted decision tree classifier that minimizes a proxy\footnote{For the empirical analysis, we estimated the weighted decision tree classifier by splitting based on the (weighted) gini impurity because the machine learning library we used (scikit-learn) did not offer the option of learning the tree by directly optimizing for the weighted misclassifcation rate.} of $WMR$ (Equation \ref{eq:wmr}) to learn $\hat{\theta}$ (Equation \ref{eq:assignment}). 

The models were learned, tuned, and evaluated using 10-fold nested cross validation, which separates the cross-validation used for hyperparameter optimization from the test folds used for evaluation~\citep{provost2013data}. All hyperparamenters were tuned to optimize their respective loss functions: the O-learner was tuned to optimize $MSE_\mu$; the E-learner was tuned to optimize $MSE_\tau$, and the A-learner was tuned to optimize $WMR$. We used the empirical measure described in Equation~\ref{eq:empirical_measure} to evaluate all policies, using $\mathbb{P}(t_i)=86.68\%$ when $t_i=0$ (control), and $\mathbb{P}(t_i)=4.44\%$ otherwise. However, for clarity, the analysis that follows compares this quantity relative to assigning the incumbent (control) system to everyone, which is the percentage increase in streams. 

\subsection{Results}

As mentioned, all metalearners eventually converge to the same treatment assignment policy if the training data is large enough. However, most firms don't have access to unlimited experimental data, or even an experimental data set as large as the one presented in this study. Thus, this analysis assesses how do the metalearners compare with various data sizes. 

\begin{figure}
    \centering
    \includegraphics[width=0.7\textwidth]{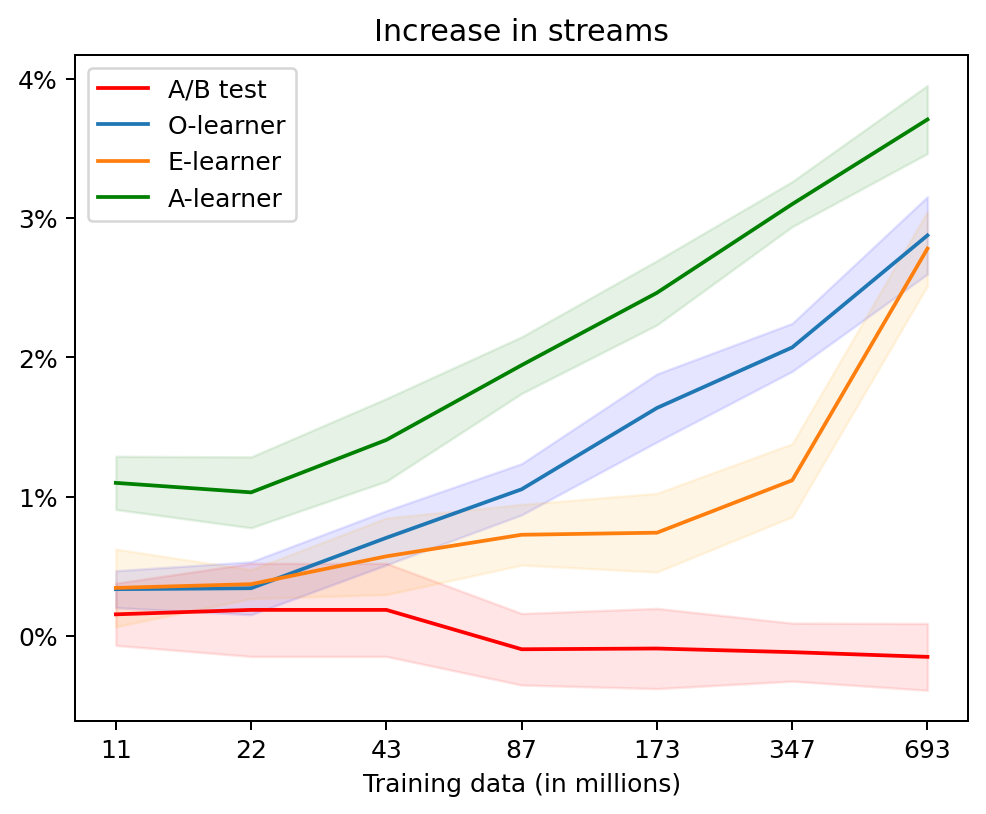}
    \caption{Treatment assignment performance by data size. All treatment assignment metalearners (blue, orange, and green) improve substantially with more data and work better than assigning everyone to the system that works best according to the A/B test (red). Also, optimizing for assignment prediction (A-learner) works better than optimizing for outcome prediction (O-learner) or causal effect prediction (E-learner).}
    \label{fig:results}
\end{figure}

Figure~\ref{fig:results} shows the performance of each metalearner (measured as the increase in streams relative to the baseline) as the size of the training data increases. The red line is the policy where the system that performs best on average is applied to everyone---this is what we would get from a standard A/B test. The other lines correspond to the treatment assignment policies estimated by the O-learner (blue), the E-learner (orange), and the A-learner (green). The areas around the lines represent 95\% confidence intervals calculated using the ten results from the cross-validations.

\subsubsection{The importance of individualized treatment assignment policies.}

The first interesting finding in Figure~\ref{fig:results} is that choosing the system that performs best on average does not significantly increase the total number of song streams.  This implies that no single `best system' for all users performs much better than the baseline (existing production system). Importantly, if we were to follow a traditional A/B test approach, we might erroneously decide to use the system currently in production for everyone, because no other system produces an increase in streams that is statistically significant at the population level. 

We show in Table~\ref{tab:assignment} the percentage of users that would be assigned to each system when the entire data are used to estimate treatment assignment policies. Recall that our analysis was conducted using nested cross-validation, so the table was built using out-of-sample treatment assignments for all users. The first row in the table shows that different systems may be selected as the `best system' (on average) depending on the cross-validation folds used to compare the systems: System $T=1$ is selected for 6 out of 10 folds, whereas System $T=2$ is selected for the other 4 folds. Combined with Figure~\ref{fig:results}, this suggests that there is no real performance difference between the systems when applied to the entire population. 

However, total streams can be increased substantially when different systems are applied to different users. Table~\ref{tab:assignment} shows that the policies that were estimated using machine learning (O-learner, E-learner, and A-learner) exhibit high heterogeneity in their treatment assignments, and these policies also perform substantially better (as shown in Figure~\ref{fig:results}) than assigning the single best playlist generation system to everyone.  To control for the fact that high heterogeneity may be the result of using different folds to estimate the models, we computed the entropy in treatment assignments for each test fold, and then obtained the average entropy across all 10 folds (i.e., column 4 in Table~\ref{tab:assignment}). As we can see, all metalearners exhibit a large entropy (i.e., high heterogeneity) in treatment assignments within the folds, resulting in substantially more streams.

\begin{table}[]
    \renewcommand{\arraystretch}{0.7} 
    \centering
    \begin{tabular}{c|c|c|c|c|c}
    \toprule
    \textbf{Policy}&\textbf{$T=0$}&\textbf{$T=1$}&\textbf{$T=2$}&\textbf{$T=3$}& Average Entropy$^*$\\
    \midrule
         Best on average$^\dagger$ &  0.0\% & 60.0\% & 40.0\% & 0.0\%& 0\\
         O-learner &  11.1\% & 25.3\% & 33.5\% & 30.1\%& 1.896\\
         E-learner &  9.3\% & 32.1\% & 29.0\% & 29.5\% & 1.879\\
         A-learner &  9.3\% & 32.0\% & 29.4\% & 29.3\%& 1.879\\
        \bottomrule
        \multicolumn{6}{l}{\footnotesize{$*$ Average entropy of treatment assignments across folds. Min is 0 and Max is 2.}} \\
        \multicolumn{6}{l}{\footnotesize{$\dagger$ Different systems perform best (on average) depending on the folds that are used}\vspace{-5pt}} \\
        \multicolumn{6}{l}{\footnotesize{to select the `best system'. Thus, not everyone is assigned to a single system when}\vspace{-5pt}}\\
        \multicolumn{6}{l}{\footnotesize{using cross-validation to evaluate and analyze the `best on average' policy.}}\\
        \hline
    \end{tabular}
    \caption{Percentage of users assigned to each treatment}
    \label{tab:assignment}
\end{table}

\subsubsection{The importance of large A/B tests.}

Another important result is that treatment assignment policies become increasingly better with more training data, illustrating the importance of conducting large A/B tests to generate unconfounded training data to learn models for personalized treatment assignments. As we mentioned in Section~\ref{sec:definition}, the estimation of causal effects by using CATEs instead of the ATE is a substantial improvement for the purposes of deciding on individual interventions. However, the estimation of accurate CATEs requires much larger data sets; otherwise the models are likely to overfit. Conducting large A/B tests alleviates this problem because the data can be partitioned into fine-grained subpopulations of users without losing as much statistical power. Correspondingly, we can fit more complex models with less overfitting. 

\subsubsection{The importance of the learning objective.}\label{sec:imp_obj}

Figure~\ref{fig:results} also shows that learning policies by optimizing treatment assignments (A-learner, green line) works better than learning policies by optimizing outcome and causal effect predictions (O-learner and E-learner, blue line and orange line respectively), thus validating our analytical findings. Importantly, this is the case \emph{even} when the training data contains more than half a billion observations. As discussed in detail, objective functions that optimize things other than treatment assignment prediction (i.e., better outcome or causal effect predictions) do not necessarily favor better treatment assignments. 

Going back to our analytical examples, we would expect each metalearner to perform best doing whatever it is optimized to do. For example, the predictive model estimated by the O-learner should perform better at predicting outcomes than the models estimated by the E-learner and the A-learner. However, as discussed in Section~\ref{sec:generality}, we cannot (in general) use causal effect models or treatment assignment models to estimate outcomes. Therefore, in order to compare the performance of the metalearners at different tasks, we adapt the models estimated by the E-learner and the A-learner in the following analysis. 

\begin{table}[]
    \renewcommand{\arraystretch}{0.7}
    \centering
    \begin{tabular}{c|c|c|c}
    \toprule
    \textbf{Metalearners}&\textbf{$MSE_\mu$}&\textbf{$MSE_\tau$}$^*$&Increase in streams$^\dagger$ \\
    \midrule
         O-learner &  \textbf{0.057} & 46.287 & 2.88\%\\
         E-learner &  0.111 & \textbf{46.276} & 2.78\%\\
         A-learner &  0.059 & 46.305 & \textbf{3.71\%}\\
        \bottomrule
        \multicolumn{4}{l}{\footnotesize{$*$ $MSE_\tau$ corresponds to the $MSE$ of the transformed outcome proposed} \vspace{-5pt}} \\
        \multicolumn{4}{l}{\footnotesize{by \cite{athey2016recursive} to estimate causal effects.}} \\
        \multicolumn{4}{l}{\footnotesize{$\dagger$ Relative to the system in production}.} \\
        \hline
    \end{tabular}
    \caption{Policy comparison at different tasks}
    \label{tab:comparison}
\end{table}

Since the metalearners in our analysis estimate tree-based models, we can generalize the E-learner and the A-learner models by using different prediction functions to aggregate the training observations at each leaf depending on the task at hand. For instance, if we want to make outcome predictions, the prediction function would consist of the average outcome of the observations in the leaf (rather than the average causal effect in the case of the E-learner model or the treatment with the largest average outcome in the case of the A-learner model). Thus, the structure of each tree model remains the same, but the prediction function at the leaf level may be adjusted to predict outcomes, causal effects, or best treatment assignments.

Table~\ref{tab:comparison} shows the performance of each metalearner at the three different tasks, evaluated using nested cross-validation on the entire data set. The tasks are predicting outcomes (a lower $MSE_\mu$ is better), predicting causal effects (a lower $MSE_\tau$ is better), and predicting treatment assignments (where the goal is to increase song streams). As expected, each metalearner is best at doing what it was optimized to do. Thus, the models with the best performance in outcome prediction (O-learner) and causal effect prediction (E-learner) are not the best models at making treatment assignments. In fact, the improvement in streams produced by the A-learner is more than 28\% larger than the improvement produced by either the O-learner or the E-learner.

\section{Discussion}

This paper groups treatment assignment methods into three general metalearners: the outcome learner (O-learners), the causal effect learner (E-learners), and the treatment assignment learner (A-learner). The grouping allows us to compare treatment assignment methods in terms of (1) their level of generality and (2) the objective function they use to learn models from data; both of these characteristics have important implications for modeling and decision making, as discussed in detail in Section~\ref{sec:definition}.

One of the major implications is that optimizing for outcome or causal effect predictions (O-learner and E-learner, respectively) is not the same as optimizing for treatment assignments (A-learner), so the latter ought to perform better in practical (non-asymptotic) settings due to its ability to exploit the bias-variance tradeoff that results from framing the treatment assignment problem as a classification task instead of a numeric prediction task. We support this claim analytically and also empirically for the real-world application of choosing, for each listener, which playlist generation algorithm to apply in order to maximize the number of song streams. Although, in theory, all metalearners should converge to the same (optimal) treatment assignment policy with unlimited data, we find that the A-learner's advantage over the other metalearners can persist even when treatment assignment policies are estimated with more than half a billion observations. 

Our study also illustrates how large A/B tests can provide substantial value for learning treatment assignment policies (rather than simply choosing the variant that performs best on average). In our application, none of the individual treatments (the different playlist generation algorithms) increases streams when applied over the whole population, but the best of the treatment assignment policies we estimated can increase the number of song streams by 3.7\%. 

\subsection{Methods for treatment assignment}\label{sec:review}

Our metalearner categorization encompasses many methods proposed in the literature for learning treatment assignment policies from data. We discuss below how the methods proposed in multiple fields of study fit into our categorization (see Table~\ref{tab:summary} for a summary).

\begin{table}
    \renewcommand{\arraystretch}{0.7} 
    \centering
    \begin{tabular}{c|l}
        \toprule
        \textbf{Field of study}&  \multicolumn{1}{c}{\textbf{\centering Discussion about approaches}} \\
        \midrule
                Econometrics & Recommendations include the O-learner and the E-learner. \\
                &Recent studies recommend the A-learner and show it should be  \\
                & preferred under certain theoretical conditions. \\
         \hline
        Uplift modeling & Comparisons and recommendations include the O-learner and the \\
        & E-learner. The E-learner is often the preferred approach, but the \\
        & O-learner has been shown to be competitive in benchmark studies. \\
         \hline
        Causal Effect Estimation & The E-learner is often recommended for treatment assignment, \\ & including in IS and marketing.\\
         \hline

        Multi-armed bandits & The O-learner is the most commonly used approach, but the \\
        & A-learner was first proposed and recommended in this field. \\
         \hline
        Treatment assignment & Some studies have noted the close similarity between treatment \\
        as classification  & assignment and classification. They recommend the A-learner. \\
         \bottomrule
    \end{tabular}
    \hspace{1cm}
    \caption{Summary of literature for treatment assignment}
    \label{tab:summary}
\end{table}

\textbf{Econometrics} has long recognized that treatment assignment is a distinct problem from the point estimation and hypothesis testing problems usually considered in the treatment effects literature~\citep{manski2004statistical,dehejia2005program,bhattacharya2012inferring,hirano2009asymptotics}. However, most of the methods proposed in the econometrics literature correspond to instances of the O-learner and the E-learner. As we discuss in Section~\ref{sec:past_comparisons}, some recent studies recommend the A-learner, showing that it is statistically efficient under a wide range of theoretical conditions \citep{kitagawa2018should,athey2021policy}.

\textbf{Uplift modeling} consists of assigning treatments based on the estimation of the incremental (causal) impact of a treatment on individuals' behaviors~\citep{lo2002true,kane2014mining}.  It has been recommended by the data mining community for targeting applications such as online advertising and customer retention~\citep{radcliffe2011real}. The uplift modeling literature typically focuses on settings where treatment assignments and outcomes are binary, so methods are usually grouped into two main categories~\citep{rzepakowski2012decision}: the two-model approach and the single-model approach (which are specific instances of the O-learner and the E-learner respectively). Most studies favor the use of the E-learner over the O-learner, but benchmark studies show that the O-learner can perform better depending on the data set~\citep{jaskowski2012uplift, olaya2020survey}. To our knowledge, none of these studies have considered the A-learner. 

\textbf{Causal Effect Estimation}.
As mentioned, most of the causal inference literature focuses on the estimation of causal effects rather than treatment assignment policies. However, the main motivation behind the use of machine learning methods for CATE estimation is often treatment assignment~\citep{athey2019machine,athey2017state}, which corresponds to the E-learner. Popular machine learning methods for CATE estimation include Bayesian additive regression trees~\citep{hill2011bayesian}, causal random forests~\citep{wager2018estimation}, and regularized causal support vector machines~\citep{imai2013estimating}. There is also a relatively large number of papers showing asymptotic properties of the E-learner for treatment assignment when an efficient or consistent estimator of CATEs is known~\citep[see][for an overview]{kitagawa2018should,athey2021policy}, but most of them do not discuss the results of deploying such systems in practice. ~\cite{dorie2019automated} provide an overview and a benchmark of several CATE estimation methods. Recent studies in the marketing and IS literature recommend such methods for treatment assignment (see Section~\ref{sec:generalization} for examples). 

\textbf{Multi-armed bandits}. Treatment assignment policies are also at the core of contextual multi-armed bandits. Models for multi-armed bandit problems may be used to learn how to make decisions in situations where the payoff of only one choice is observed~\citep{beygelzimer2009offset,dudik2011doubly}. Such methods have been used to make automated decisions about online news recommendations to maximize clicks~\citep{li2010contextual}, for example. It is precisely in this stream of research that it was first noted that the treatment assignment problem (as defined in Section~\ref{sec:problem}) is mathematically equivalent to a weighted classification problem~\citep{zadrozny2003policy,beygelzimer2009offset}, leading to the suggestion of the A-learner. Nonetheless, the O-learner has also been recommended for multi-armed bandit algorithms---LinUCB being a well-known example~\citep{li2010contextual}---and remains the most common approach in multi-armed bandit problems.
 
 An important distinction between our setting and the multi-armed bandit problem is that the goal in bandit problems is to learn a treatment assignment policy while actively making treatment assignment decisions for incoming subjects. Therefore, there is an exploration/exploitation dilemma that plays an important role in the decision making procedure, whereas in our case the decision-maker cannot re-estimate the treatment assignment policy after making each decision.\footnote{Few firms have the ability to deploy full-blown online machine learning systems that can manage the exploration/exploitation tradeoff dynamically.  It is much more common to deploy the learned models/prediction systems than the machine learning systems themselves.} Our setting is also referred to as ``offline learning'' in this literature~\citep{beygelzimer2009offset}.
 
 \textbf{Treatment assignment from a classification perspective}. \cite{zhang2012estimating} proposed a generalized classification framework to show how several estimators of optimal treatment regimes can be represented as special cases of weighted classification within their framework; these estimators are essentially instances of the A-learner. This framework defines weights in terms of a contrast function that may represent outcomes \citep[as in][]{zhang2012estimating,kitagawa2018should}, causal effects \citep[as in][]{athey2021policy}, or some other business-oriented importance weight~\citep[as in][]{lemmens2020managing}. In the personalized medicine literature, the A-learner is also referred as ``outcome weighted learning"~\citep{zhao2012estimating,chakraborty2013statistical}. 

\subsection{Prior comparisons of the metalearners}\label{sec:past_comparisons}

To our knowledge, no prior study has compared the three metalearners as defined in our study, either analytically or on a real-world application at scale, but there are some partial exceptions.

\cite{schuler2018comparison} propose a framework to compare and select models based on their ability to predict outcomes, causal effects, and optimal treatments, but they do not consider how learning models based on these three criteria may affect treatment assignment performance. Interestingly, they show through simulations that selecting models based on their ability to predict causal effects generally leads to better treatment assignments than selecting models based on their ability to predict outcomes or optimal treatments (hence suggesting that the E-learner is a good candidate for treatment assignment). However, their comparisons do not include the A-learner.

\cite{beygelzimer2009offset} provide a theoretical regret analysis for multi-armed bandit problems showing that, for a given family of regressors (e.g., decision trees), the A-learner has a smaller lower bound regret than the O-learner. These analytical results are supported by experiments on multi-class benchmark data sets that were repurposed to simulate potential outcomes, showing the A-learner as a superior alternative than the O-learner. 

Other recent studies have also made theoretical developments that are in line with our study and argue in favor of the A-learner. \cite{kitagawa2018should} investigate the statistical performance of the A-learner (to which they refer as Empirical Welfare Maximization) in terms of its uniform convergence rate of regret. They show that in settings where propensity scores are known (e.g., when data are acquired through A/B tests), the A-learner attains minimax optimal rates in finite samples over various classes of feasible data distributions.~\cite{athey2021policy} extend these results to show the asymptotical efficiency of the A-learner when estimating policies from observational data, so they address cases where (1)~propensity scores are not known or (2) there is endogeneity (so unconfoundedness is not met) but effects can be estimated using instrumental variables.

Our study builds on this past work in several important ways. First, we elaborate on the advantages and disadvantages of the A-learner with respect to other metalearners commonly proposed in the treatment assignment literature. Second, we show that the A-learner can outperform other metalearners as a result of optimizing the bias-variance tradeoff with respect to decision making errors~(i.e., treatment assignment) rather than conventional prediction errors in outcomes or causal effects. Third, we empirically demonstrate the advantages of the A-learner for the deployment of content selection systems in the context of music streaming.

Other studies in the uplift modeling literature have also compared the O-learner and the E-learner~\citep{jaskowski2012uplift,olaya2020survey}. Notably, \cite{olaya2020survey} conducted an extensive benchmark study comparing the performance of thirteen uplift modeling methods that are instances of the O-learner and the E-learner as defined by our categorization. The comparisons included data sets across many domains of interest, including marketing, political behavior, and clinical trials. They found that none of the evaluated techniques consistently outperforms the other techniques. Although the uplift modeling literature generally favors the E-learner over the O-learner, the consensus is that choosing between the O-learner and the E-learner should be an empirical undertaking because the O-learner sometimes beats the E-learner~\citep{jaskowski2012uplift,olaya2020survey}. As discussed next, this may also occur when choosing between the A-learner and the other metalearners. 

\subsection{Generalizability of the results}\label{sec:generalization}
Our results do not imply that the A-learner will necessarily outperform the other metalearners in other settings. Results may change depending on the data generating process, the available features, the size of the data, and the machine learning algorithm used by the metalearners (see Appendix~\ref{app:extended} for an extended empirical analysis that considers some of these factors). However, our argument is that, in general, the A-learner should be a strong (if not the strongest) contender. 

This point is important because several recent papers in the marketing and the IS literature employ machine learning to estimate causal effects that are then used for treatment assignment~\citep{mcfowland2021prescriptive,ascarza2018retention,yang2020targeting,miller2020personalized}; these methods would be considered O-learners or E-learners under our categorization. As an exception,~\cite{lemmens2020managing} propose a profit-based loss function that uses a weighting scheme that could be cast as an instance of an A-learner; they show that their approach performs better than the O-learner when managing churn to maximize profits. However, among practitioners and scholars who consider the causal impact of treatments when making treatment assignments,\footnote{As discussed in~\cite{ascarza2018retention}, it is not uncommon for practitioners to estimate treatment assignment policies from data without any sort of causal modeling.~\cite{fernandez2019causal} give a modeling justification for this.} O-learners and E-learners are by far the most common approaches for treatment assignment in the marketing and IS literature, and many other fields of study. 

\subsection{Limitations and future research}

One important assumption in our study is that the training data is generated from A/B tests. For cases where this assumption is not met, the A-learner can be used to leverage observational data where causal effects can be identified using a variety of strategies, including
selection on observable features and instrumental variables~\citep{athey2021policy}. In a similar vein, \cite{fernandez2019observational} assessed the impact of confounding bias on treatment assignments when causal effects cannot be identified from data, but it is unclear from their study how the three metalearners would compare to each other in settings where confounding is prevalent. This presents a promising direction for future research.

The second major assumption in this work is that all decisions are independent. However, this is unlikely to be the case when multiple treatments are assigned to the same individuals over time. Treatment assignment policies for such settings are also known as dynamic treatment regimes~\citep{murphy2003optimal,chakraborty2014dynamic}. We are unaware of any A-learner methods specifically designed to model dependent decisions, making this a natural next step to extend this line of work. 

Another related setting that our study does not consider is that of constrained optimization, such as when treatment assignments are made under budget constraints. Under such circumstances, it is no longer necessarily the case that assigning individuals to the treatment with the most beneficial outcome is the optimal thing to do.~\cite{mcfowland2021prescriptive} propose a framework to maximize the overall utility of treatment assignments under budget constraints by using machine learning for predicting costs and causal effects, and then optimizing an integer linear program that converts predictions into treatment assignments. This is also known in the operations literature as a predict-then-optimize framework~\citep{elmachtoub2021smart}. That is, machine learning is used first to predict unknown input parameters of an optimization problem, and then decisions are made by solving the optimization problem using the predicted parameters. 

An implication from our study is that the assignments should be better when the objective function in the machine learning measures errors in the assignments induced by the predicted input parameters, as opposed to errors in the prediction of the input parameters themselves. Such methods have been applied in the operations literature for a variety of optimization problems with linear objectives~\citep{elmachtoub2021smart,elmachtoub2020decision}. Future research could also explore and apply these ideas in the context of treatment assignment problems, as suggested in~\cite{fernandez2022causal}.

Finally, we hope that this study will encourage other researchers to further explore other parallels between policy estimation and ``traditional'' predictive modeling, besides the classification analogy discussed in this paper. For example, policy estimation (as defined in this paper) is somewhat related to learning to rank (LTR)---as discussed in the information retrieval literature~\cite[e.g.,][]{burges2005learning}. More specifically, for each user, we could rank a set of actions according to their potential outcomes and choose the action at the top of the ranking. Of course, LTR is not the same type of task because, in our setting, we only care about the difference in outcomes between the action we choose and the action at the top of the ranking (not the entire ranking). Nonetheless, LTR methods could potentially be adapted to address treatment assignment problems, perhaps in a similar fashion to the weighted-classification transformation discussed in this paper.

%\bibliographystyle{ACM-Reference-Format}
%\bibliography{bibfile}

% Acknowledgments here
\ACKNOWLEDGMENT{%
% Enter the text of acknowledgments here
}% Leave this (end of acknowledgment)

\begin{APPENDICES}

\section{Proof that Equation~\ref{eq:empirical_measure} is an unbiased estimator of Equation~\ref{eq:max_measure}}\label{app:proof1}

We present here a simplified proof that Equation~\ref{eq:empirical_measure} is an unbiased estimator of Equation~\ref{eq:max_measure} (see~\cite{li2010contextual} for a detailed proof):
\begin{equation*}
\begin{split}
    \mathbb{E}_{X,Y,T}\Bigg[\frac{1}{n}\sum_{i=1}^n \textbf{1}(\hat{a}(X)=T)\frac{Y}{\mathbb{P}(T)}\Bigg]&=\mathbb{E}_{X,Y,T}\Bigg[\textbf{1}(\hat{a}(X)=T)\frac{Y}{\mathbb{P}(T)}\Bigg]\\
    &=\sum_{\forall j\in T}\mathbb{E}_{X,Y}\Bigg[\textbf{1}(\hat{a}(X)=j)\frac{Y}{\mathbb{P}(T=j)}\Big|T=j\Bigg]\mathbb{P}(T=j) \\
    &=\sum_{\forall j\in T}\mathbb{E}_{X,Y}[\textbf{1}(\hat{a}(X)=j)Y|T=j], \\
\end{split}
\end{equation*}
and given the unconfoundedness assumption:
\begin{equation*}
\begin{split}
&=\sum_{\forall j\in T}\mathbb{E}_{X,Y(j)}[\textbf{1}(\hat{a}(X)=j)Y(j)] \\
&=\mathbb{E}_{X,\{Y(j):j\in T\}}\Bigg[\sum_{\forall j\in T}\textbf{1}(\hat{a}(X)=j)Y(j)\Bigg] \\
&=\mathbb{E}_{Y(\hat{a}(X))}[Y(\hat{a}(X))] \\    
\end{split}    
\end{equation*}

\begin{multline*}
\\ 
\qed
\end{multline*}

\section{Proof that minimizing weighted misclassification rate (Equation~\ref{eq:wmr}) is equivalent to minimizing expected regret (Equation~\ref{eq:min_measure})}\label{app:proof2}

We present here a simplified proof that minimizing Equation~\ref{eq:wmr} is equivalent to minimizing Equation~\ref{eq:min_measure} (see~\cite{beygelzimer2009offset} for more details):
\begin{equation*}\label{eq:wmr_proof}
\begin{split}
    \argmin_{\hat{a}}~WMR(\hat{a})&=\argmin_{\hat{a}}~ \mathbb{E}_{X,Y,T}\Bigg[\textbf{1}(\hat{a}(X)\neq T)\frac{Y}{\mathbb{P}(T)}\Bigg] \\
    &= \argmin_{\hat{a}}~ \mathbb{E}_{Y, T}\Bigg[\frac{Y}{\mathbb{P}(T)}\Bigg] - \mathbb{E}_{X,Y,T}\Bigg[\textbf{1}(\hat{a}(X)=T)\frac{Y}{\mathbb{P}(T)}\Bigg]\\
    &= \argmin_{\hat{a}}~-\mathbb{E}_{X,Y,T}\Bigg[\textbf{1}(\hat{a}(X)=T)\frac{Y}{\mathbb{P}(T)}\Bigg],
\end{split}
\end{equation*}
and from the proof in Appendix~\ref{app:proof1}, it follows that:
\begin{equation*}
\begin{split}
        &= \argmin_{\hat{a}} - \mathbb{E}_{Y(\hat{a}(X))}[Y(\hat{a}(X))] \\
        &= \argmin_{\hat{a}}\mathbb{E}_{Y(a^*(X)),Y(\hat{a}(X))}[Y(a^*(X))-Y(\hat{a}(X))]  \\
        &= \argmin_{\hat{a}}~ \text{Regret}(\hat{a})
\end{split}
\end{equation*}
\begin{multline*}
\\ 
\qed
\end{multline*}

\section{Data generating process of the simulated example in Section~\ref{sec:ap_desc}}~\label{app:data}

Below are the formulations for all variables in the data generation process, including the potential outcome when treated, $Y(1)$; the potential outcome when untreated, $Y(0)$; and the feature used to make intervention decisions, $X$. The potential outcomes are normally distributed and the feature is uniformly distributed:
\begin{align}
X \sim&~\mathcal{U}_{[0,1]} \\
Y(1) \sim&~\mathcal{N}(\mu_1(X), 0.2) \\
Y(0) \sim&~\mathcal{N}(\mu_0(X), 0.2) \\
\mu_1(x) =& \frac{1}{1+e^{2-5x}} + 0.5 \\
\mu_0(x) =& \frac{0.1}{x+0.4} + 0.5x^2 + 0.5
\end{align}

The data we generated consists of 50 treated observations and 50 untreated observations. 

\section{Comparison to conventional treatment assignment problems}\label{app:diffs}
 
There are a few ways in which the deployment of content selection systems as treatments is different from conventional treatment assignment problems. First, one system can be a better (or worse) treatment than another system only to the extent that it generates different content (i.e., playlists that have different songs or rank songs differently). If the content produced for a given user does not vary across systems, then all systems are essentially the same treatment for that user, and the user's engagement cannot be improved by deploying a different variant than the system currently in production. 

Therefore, helpful features for treatment assignment are not only the ones that describe user behavior---as in conventional treatment assignment problems---but also the ones that describe how the system's content could be affected (relative to other systems). Going back to our example with new and experienced users, if System B relies on a more complex machine learning system than System A, then we may expect System B to be more susceptible to cold start problems, and as a result user tenure could be a good candidate feature to decide whether to assign one system or the other. Additionally, user behavior could also play a role when deciding what system to assign. For example, if System B generally produces more niche playlists than System A (e.g., the former considers songs in languages other than English whereas the latter does not), we may also want to consider features that describe users that may appreciate such playlists (e.g., the country where users live). 

Second, when deploying systems as treatments, features may need to be selected based on the feasibility of actually using them to make system assignments. More specifically, implementation challenges may prevent us from using features that are helpful to determine whether one system is better than another in a particular context. For instance, even though ``time active in session'' may be a useful feature to decide whether to assign a system that produces playlists based on the last songs the user has played, using this feature effectively would require the ability to make system assignments throughout the user session, which may be a non-trivial technical capability. On the other hand, features that change less frequently (e.g., type of subscription, country of residence) would allow making system assignments even before users log in. 

Finally, treatment assignment policies in our setting are meant to be used to make large-scale decisions automatically; millions of playlists are generated by Spotify every day, each in a fraction of a second. This presents a salient contrast to other settings where treatment assignment policies are intended to support and improve human decision-making, such as in medical prescriptions and public policy. In such settings, transparency is critical and the rationale behind each decision must be clearly understood.

Understanding the rationale behind treatment assignments can also be important in our context. For example, deeper understanding of the most important factors for system assignment may help to inform the development of new playlist generation systems (i.e., new `treatments') that may further improve engagement. However, our study focuses on how to learn a policy to identify the system that leads to the highest expected number of song streams for each user, not on understanding what are the factors that result in one system being better than another. 

\section{Extended analysis}\label{app:extended}

In this extended empirical analysis, we consider how (1) the available features and (2) the machine learning procedure can influence the preference of one metalearner over the others. 

\subsection{The importance of the available features}

The usefulness of estimating treatment assignment policies with machine learning depends to a large extent on the features that are available to classify individuals into different treatments. If the features are not informative of preferred treatments, then there is no advantage over using a standard A/B test to determine (and assign) the alternative that works best on average. However, with more features, it also becomes harder to estimate \emph{optimal} treatment assignments (from a regret minimization perspective) because the optimal policy also becomes more complex. Thus, typically, the more features there are, the more training data it takes for the metalearners to converge to the optimal treatment assignment policy.

\begin{figure}
    \centering
    \includegraphics[width=0.7\textwidth]{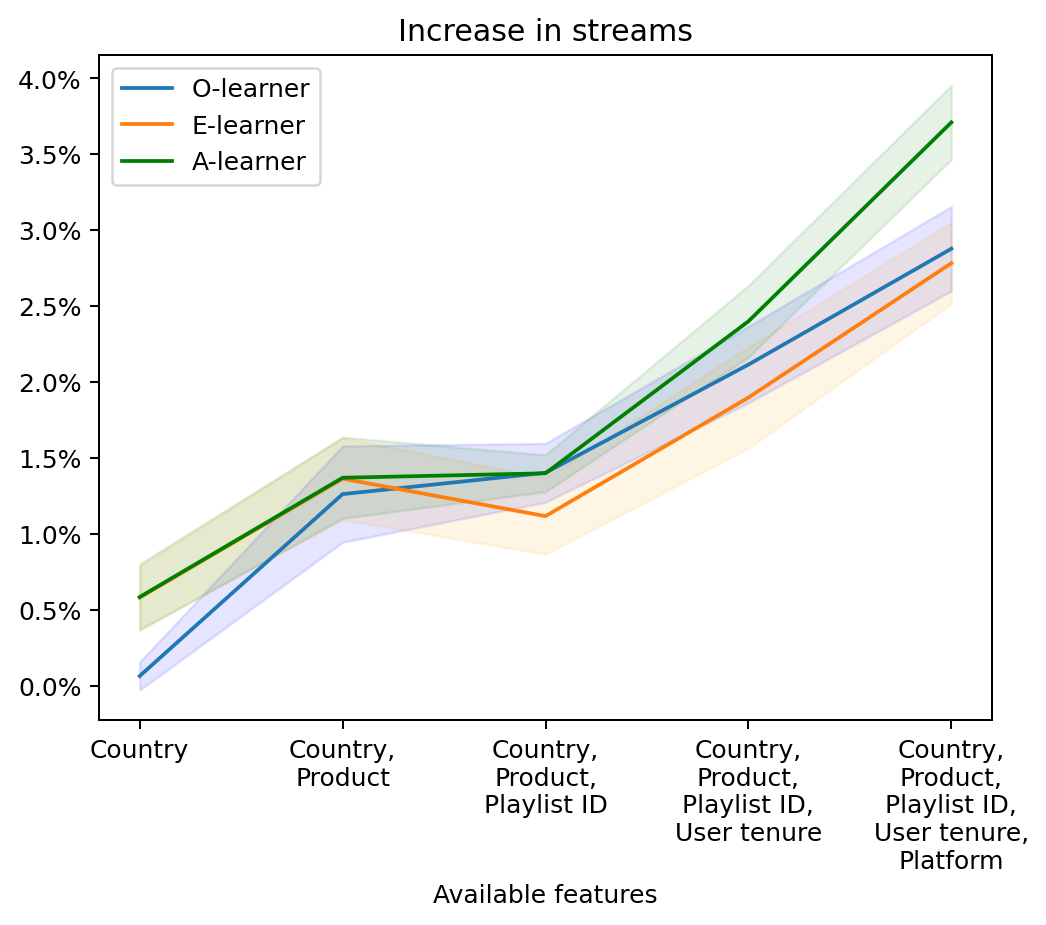}
    \caption{Treatment assignment performance by number of features. All treatment assignment metalearners improve substantially with more features. The A-learner is more likely to outperform the other metalearners when the number of features is larger.}
    \label{fig:features}
\end{figure}

Figure~\ref{fig:features} shows the results of deploying treatment assignment policies that were estimated with an increasing number of features. Features are ordered from the one with the most categorical values to the one with the least categorical values. As before, the lines correspond to each of the three metalearners---O-learner (blue), E-learner (orange) and A-learner (green)---and the areas around the lines represent 95\% confidence intervals. As expected, all metalearners perform better with more features because they have more information about the individual to make assignments. Furthermore, the metalearners have a similar performance when there are few features because there is enough training data for them to converge to the same policy. For example, the A-learner performs the same as the O-learner when there are 3 features and the same as the E-learner when there are one or two features. As mentioned, this is expected because machine learning algorithms converge faster to best-in-class models when there are fewer features. Trees are universal approximators, so they lead to the same (optimal) treatment assignment policy given enough training data. 

On the other hand, as the number of features increases, the A-learner becomes better than the other metalearners. As discussed in Section~\ref{sec:obj_discussion}, the A-learner focuses on minimizing prediction errors that negatively affect decision making and ignores all other types of errors. As a result, it does a better job at exploring the feature space and discriminating individuals based on their preferred treatment assignment (rather than based on causal effects or outcomes).

\subsection{The importance of the machine learning algorithm.} \label{sec:algorithm}

In this extended analysis, we show how the choice of the base learner may also affect the comparison of the metalearners. We consider multiple base learners here: the tree-induction algorithms discussed in Section~\ref{sec:learning_eval}, random forest, and generalized linear regression. Random forest was implemented using ensembles of the tree-based methods described in Section~\ref{sec:learning_eval}. In the case of the generalized linear regression models, for the O-learner, we used a linear regression for each treatment condition. For the E-learner, we used a linear regression on the transformed variable proposed by~\cite{athey2016recursive} for each of the 3 non-control systems. Finally, for the A-learner, we used a weighted logistic regression. 

Figure~\ref{fig:alg_per} shows the performance of the methods we considered. The black lines represent 95\% confidence intervals. These confidence intervals describe deviations in the \emph{individual} performance of each method, but they are not appropriate to compare the metalearners because they do not account for the correlation in the deviations (e.g., when one method works particularly well in a fold, the other methods will also tend to work particularly well). Importantly, this figure shows that the best performing method corresponds to the A-learner in our main analysis (i.e., the one that uses tree induction as a base learner). 

\begin{figure}
    \centering
    \includegraphics[width=0.7\textwidth]{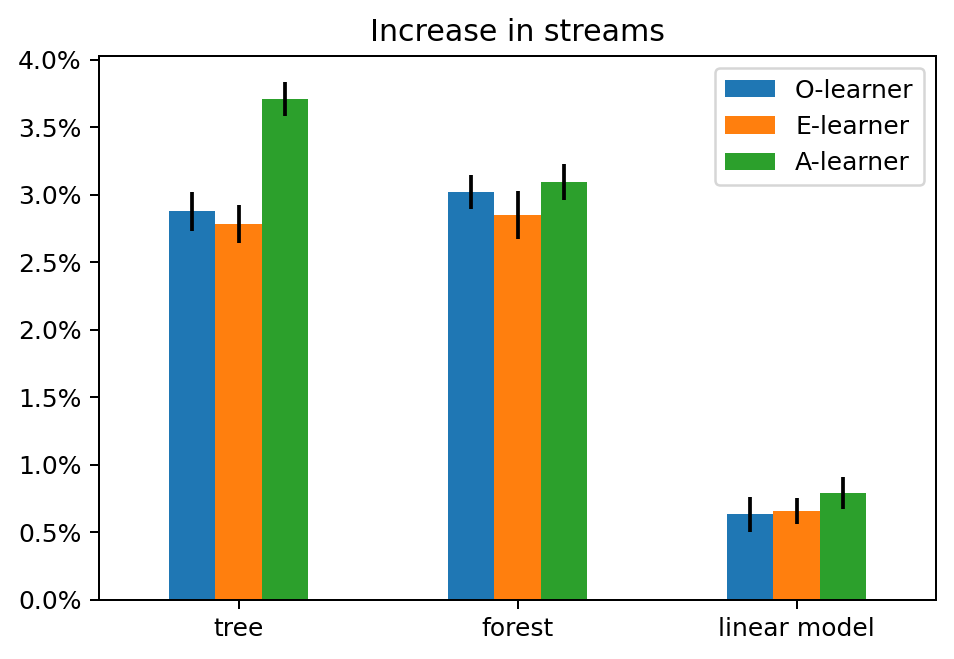}
    \caption{Treatment assignment performance by type of metalearner and base learner. The A-learner with the tree base learner works best.}
    \label{fig:alg_per}
\end{figure}

Figure~\ref{fig:alg_imp} appropriately shows the difference in performance between metalearners. For example, ``from O-learner to A-learner'' corresponds to the the difference in performance between the A-learner and the O-learner (i.e., the difference between green bars and blue bars in Figure~\ref{fig:alg_per}). The black lines represent 95\% confidence intervals. The A-learner is the best performing metalearner in all cases, but the improvement is not as large when random forests and linear models are used. Thus, as discussed in Section~\ref{sec:generalization}, results can be affected by the base learner.

\begin{figure}
    \centering
    \includegraphics[width=0.7\textwidth]{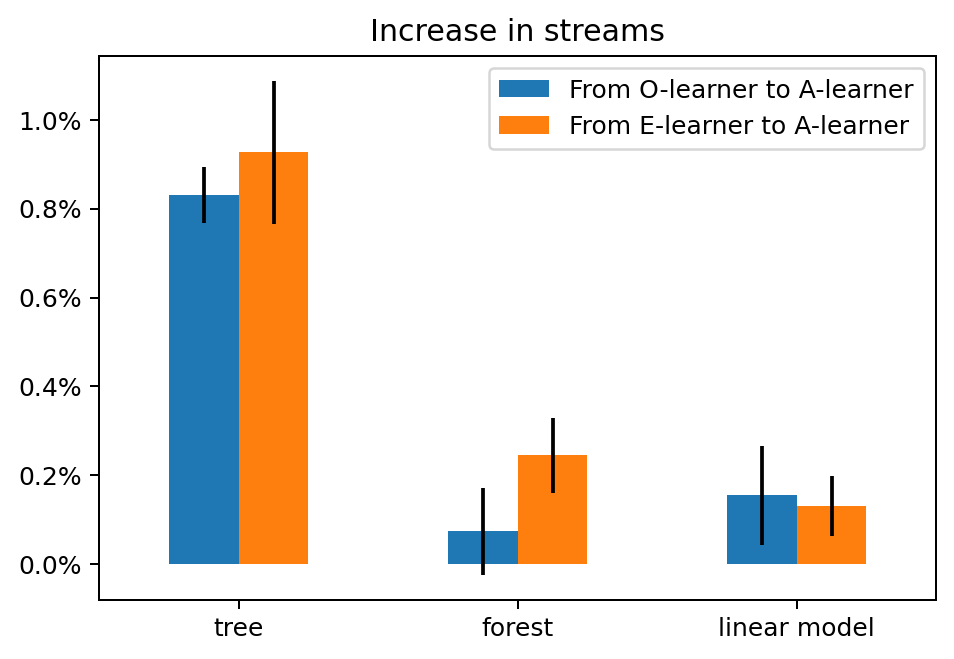}
    \caption{Increase in streams when the A-learner is used instead of the other meta-learners. The A-learner generally works better, but the improvement is moderated by the base learner.}
    \label{fig:alg_imp}
\end{figure}

There are multiple points worth discussing here. First, how the base learner affects the comparison of the metalearners can vary from one data set to another. For example, \cite{olaya2020survey} compare various instances of the E-learner and the O-learner across multiple data sets using random forest and logistic regression as base learners. They find that none of the evaluated techniques consistently outperforms the other techniques, which implies that choosing among metalearners (and base learners) should be an empirical undertaking. The results we present here just show that the choice of the base learner can affect how the metalearners compare to each other; they do not imply that, in general, the difference between metalearners should be smaller or larger for any of the base learners under consideration.

Second, recall that, when the base learner is a consistent estimator, all metalearners estimate the same (optimal) treatment assignment policy with large enough data. This implies that, in general, differences in the estimated policies will be larger with smaller data. Therefore, given that most firms cannot conduct A/B tests as large as the one in this study (770 million observations), we should expect to observe larger differences in performance between metalearners in other practical settings.

Third, one possible explanation for the results in Figure~\ref{fig:alg_imp} is that tree models suffer from higher variance than random forests and linear models, and as a result, differences between the metalearners will tend to be larger when trees are used as base learners. Nevertheless, the distinction between metalearners can also be important when low-variance models are used (e.g., linear models) because bias also plays an important role in the learning procedure. As discussed in Section~\ref{sec:obj_discussion}, the A-learner focuses on minimizing bias that negatively affects decision making, whereas the O-learner and the E-learner focus on minimizing bias that negatively affects outcome and causal effect predictions. Therefore, in general, the A-learner will converge to better policies than the other metalearners when biased base learners are used.~\cite{elmachtoub2021smart} provide evidence of this for non-causal decision making.

%\begin{figure}
%    \centering
%    \includegraphics[width=0.7\textwidth]{performance_by_algorithm.png}
%    \caption{Treatment assignment performance by type of model. Non-linear models (tree and forest) lead to better treatment assignment policies than linear models for all metalearners.}
%    \label{fig:alg_per}
%\end{figure}

%Figure~\ref{fig:alg_per} shows the treatment assignment performance of the three metalearners (OL, EL, and AL) when implemented with the three types of machine learning algorithms under consideration (trees, forests, and linear models). The black lines represent 95\% confidence intervals. Interestingly, Figure~\ref{fig:alg_per} shows that the trees and forests perform substantially better than the linear models, suggesting that modeling interactions between features is quite important in our setting.

\end{APPENDICES}

\bibliographystyle{ACM-Reference-Format}
\bibliography{bibfile}

% Appendix here
% Options are (1) APPENDIX (with or without general title) or 
%             (2) APPENDICES (if it has more than one unrelated sections)
% Outcomment the appropriate case if necessary
%
% \begin{APPENDIX}{<Title of the Appendix>}
% \end{APPENDIX}
%
%   or 
%
% \begin{APPENDICES}
% \section{<Title of Section A>}
% \section{<Title of Section B>}
% etc
% \end{APPENDICES}

% References here (outcomment the appropriate case) 

% CASE 1: BiBTeX used to constantly update the references 
%   (while the paper is being written).
%\bibliographystyle{informs2014} % outcomment this and next line in Case 1
%\bibliography{<your bib file(s)>} % if more than one, comma separated

% CASE 2: BiBTeX used to generate mypaper.bbl (to be further fine tuned)
%\input{mypaper.bbl} % outcomment this line in Case 2

\end{document}